\documentclass{elsarticle}
\journal{Journal of Computational Physics}

\usepackage{algorithm}
\usepackage[noend]{algpseudocode}
\usepackage{tikz}
\usetikzlibrary{arrows.meta,positioning,fit,backgrounds,calc}
\usepackage{amsmath}
\usepackage{bm}
\usepackage{bbm}
\usepackage{hyperref}
\usepackage{cleveref}
\usepackage{color}
\usepackage[letterpaper,top=1in,bottom=1in,left=1in,right=1in]{geometry}
\usepackage{lineno}
\usepackage{mathrsfs}
\usepackage{subcaption}
\usepackage{graphicx}
\usepackage{grffile}
\usepackage{amsthm}
\usepackage{enumitem}
\usepackage{amssymb}
\usepackage[group-separator={,}]{siunitx}
\usepackage{booktabs}
\usepackage{tabularx}
\usepackage{siunitx}
\usepackage[colorinlistoftodos,prependcaption,textsize=small]{todonotes}
\usepackage{placeins}

\newif\ifdraft
\drafttrue
\ifdraft
\newcommand{\notel}[1]{\todo[inline,color=blue!40]{#1}}
\newcommand{\notem}[1]{\todo[inline,color=green!40]{#1}}
\newcommand{\noteh}[1]{\todo[inline,color=red!40]{#1}}
\else
\newcommand{\notel}[1]{}
\newcommand{\notem}[1]{}
\newcommand{\noteh}[1]{}
\fi

\begin{document}
\begin{frontmatter}
\title{A Scalable Time-Based Molecular Dynamics Approach for \\ Simulating Single-Bubble Sonoluminescence}
\author[cs,isis]{Shihan Cheng}
\cortext[cor1]{Corresponding author}
\ead{shihan.cheng@vanderbilt.edu}
\author[cs,isis]{David A.\ B.\ Hyde\corref{cor1}}
\ead{david.hyde.1@vanderbilt.edu}
\address[cs]{Department of Computer Science, Vanderbilt University, 1400 18\textsuperscript{th} Avenue South, Nashville, TN 37212-2846, USA}
\address[isis]{Institute for Software Integrated Systems, Vanderbilt University, 1025 16\textsuperscript{th} Avenue South, Nashville, TN 37212-2328, USA}
\date{}

\begin{abstract}
We present a scalable time-based molecular dynamics (TBMD) framework for simulating single-bubble sonoluminescence within a hybrid continuum--MD formulation.
Unlike prior event-based approaches, which model gas dynamics through instantaneous hard-sphere collisions, the present method integrates continuous Lennard-Jones and damped shifted force Coulomb interactions at each timestep, enabling self-consistent tracking of ionization state and long-range electrostatics throughout the collapse.
To bridge the gap between the physical particle count ($N_\mathrm{real}\sim 10^{10}$) and computationally tractable ensemble sizes, we introduce an ensemble particle (EP) scaling formalism that preserves temperature, pressure, and ionization statistics while reducing the simulated particle count by up to four orders of magnitude.
Applying the framework to argon under standard single-bubble sonoluminescence driving conditions, we perform a systematic sweep over the ionization model and thermal accommodation coefficient $\alpha_t$, with ensemble sizes up to $N_\mathrm{ensem} = 10^8$ particles.
The results establish that ionization is the dominant regulator of peak temperature, reducing $T_\mathrm{max}$ by approximately a factor of two relative to the non-ionizing baseline, while $\alpha_t$ primarily controls the spatially averaged temperature at the collapse minimum.
Scalar observables at $N_\mathrm{ensem} = 10^8$, including peak temperature, minimum bubble radius, and maximum wall velocity, are assessed against prior studies to help validate the EP scaling formalism and our hybrid continuum--MD framework.
\end{abstract}

\begin{keyword}
Sonoluminescence, molecular dynamics, ensemble-particle scaling, ionization, Keller--Miksis equation
\end{keyword}

\end{frontmatter}

%\linenumbers

\section{Introduction}
\label{sec:introduction}

Sonoluminescence (SL) is the conversion of acoustic energy into light, occurring for example when a gas bubble suspended in liquid is driven by ultrasound to undergo periodic cavitation~\cite{dabh1}.
During collapse, the bubble wall accelerates to speeds approaching the local speed of sound in the surrounding liquid, compressing the enclosed gas to transient states of extreme temperature and pressure~\cite{dabh7}.
The phenomenon occurs naturally in biological systems, most notably in the cavitation produced by the snapping shrimp~\cite{dabh2}, and has attracted sustained interest for its potential medical applications, including in ultrasound-assisted lipectomy~\cite{dabh3}, acoustically activated drug delivery~\cite{dabh5}, and cavitation assessment in prosthetic heart valves~\cite{dabh4}.
%and the study of hydrodynamic cavitation in naval contexts~\cite{dabh6}.

Despite decades of theoretical and experimental investigation, the microphysics of the collapse in single-bubble sonoluminescence (SBSL) remains incompletely understood~\cite{dabh7,barber1997defining}.
The relevant events unfold over nanosecond time scales at microscopic spatial scales, making experimental characterization of the bubble's interior thermodynamic state during collapse fundamentally difficult.
Molecular dynamics (MD) simulation provides a feasible alternative for studying SBSL, with an ability to use minuscule time steps and simulate individual particle trajectories. 
%throughout the collapse without requiring closure assumptions for the equation of state or transport coefficients.
Prior simulation studies of SL~\cite{dabh8,ruuth2002molecular,dabh10,dabh11,dabh12,dabh13,kim2007molecular} have relied primarily on event-based hard-sphere MD implementations, which are inherently sequential and difficult to parallelize at large scale; to the authors' knowledge, the largest such MD study reached $10^7$ ensemble particles~\cite{dabh10}, which is deficient compared to the roughly $10^{10}$ molecules involved in representative real-world SBSL experiments.
The hard-sphere framework also provides no natural path toward including long-range Coulomb interactions between the charged carriers produced during ionization, an omission that becomes physically consequential at the extreme temperatures reached near the collapse minimum.

This work introduces the first time-based molecular dynamics (TBMD) formulation of SBSL.
Rather than resolving particle interactions as instantaneous collision events, TBMD integrates continuous equations of motion at each timestep, allowing the interaction model to combine a soft-sphere Lennard--Jones potential with a damped shifted force (DSF) Coulomb interaction~\cite{fennell2006ewald,10.1063/1.478738} between ionized species.
Implemented within the LAMMPS framework~\cite{thompson2022lammps} with custom extensions for ionization, gas--wall energy exchange, and Keller--Miksis wall coupling, our proposed method scales to at least $N_{\mathrm{ensem}} = 10^8$ ensemble particles, a tenfold increase over the prior state of the art.
%, enabling the first systematic convergence study of particle-count dependence in SL simulation.
We apply the framework to a parameter sweep of the thermal accommodation coefficient $\alpha_t$ and ionization model, quantifying how each modeling choice affects peak temperature, wall dynamics, and light emission, and we validate our results against the EBMD reference dataset of Schanz et al.~\cite{dabh10}.
Source code is released under the GPL-2.0 license.\footnote{\url{https://github.com/csh-apprentice/lammps-tbsl}}

The remainder of the paper is organized as follows.
Section~\ref{sec:related} reviews prior continuum and MD studies of SL.
Section~\ref{sec:theory} describes the continuum Keller--Miksis framework used to drive the bubble wall in our simulations.
Section~\ref{sec:md} presents the TBMD model, including interaction potentials, ionization algorithm, and gas--wall coupling.
Section~\ref{sec:ep_scaling} derives novel scaling rules for accurately setting physical quantities on ensemble particles.
With all this in hand, Section~\ref{sec:sim_proc} describes our simulation setup and diagnostics, and Section~\ref{sec:results} presents a variety of numerical results.
Section~\ref{sec:convergence} examines convergence with ensemble size to assess how many simulation particles are needed for reliable simulation results.
We conclude and discuss future directions in Section~\ref{sec:conclusions}.

\section{Related Work}
\label{sec:related}

Single-bubble sonoluminescence involves tightly coupled fluid dynamics, thermodynamics, and atomic-scale physics spanning at least twelve orders of magnitude in time.
Prior numerical work has therefore developed along two largely parallel streams: continuum partial differential equation (PDE) models that resolve the macroscopic bubble dynamics and temperature field, and molecular dynamics simulations that track discrete particle dynamics near collapse.
We review each of these research streams and then briefly discuss the MD cyberinfrastructure that enables the present work.

% -----------------------------------------------------------------------
\subsection{Continuum Modeling of Sonoluminescent Bubble Collapse}
% -----------------------------------------------------------------------

The foundation of all continuum SL modeling is an equation of motion for the spherical bubble wall.
In its simplest form, this is the Rayleigh--Plesset (RP) equation, which balances inertial, viscous, and surface tension forces on an incompressible liquid shell~\cite{prosperetti1988nonlinear}.
An early compressible-liquid formulation was given by Gilmore~\cite{gilmore1952growth}, who applied the Kirkwood--Bethe hypothesis to reduce the flow equations around the bubble to a single ordinary differential equation for the wall velocity valid into the supersonic regime.
At the extreme wall speeds reached during SL collapse, where the wall Mach number is not negligible, a compressible extension is required; the Keller--Miksis (KM) equation~\cite{keller1980bubble} incorporates acoustic radiation damping and remains accurate into the transonic regime.
We adopt the KM equation as the bubble-motion model in this work.
Broad reviews of bubble oscillation dynamics and the continuum SL modeling landscape are provided by Lauterborn and Kurz~\cite{lauterborn2010physics} and Brenner, Hilgenfeldt, and Lohse~\cite{brenner2002single}.

Coupling the bubble radius equation to the thermal and pressure state of the gas interior requires additional modeling.
Kwak and Yang~\cite{kwak1995aspect} and Kwak and Na~\cite{kwak1997physical} developed a self-contained hybrid PDE model that pairs the KM equation with a thermal boundary layer of thickness $\delta(t)$ in the surrounding liquid and derives closed-form temperature profiles under both uniform and nonuniform pressure assumptions.
This Kwak model forms the theoretical basis of the continuum reference solution used throughout our results; a description of the numerical procedure for solving the coupled ODE/PDE system is given in Section~\ref{sec:numerical_solver}.
A different continuum approach by Vuong and Szeri~\cite{vuong1996sonoluminescence} solves the Navier--Stokes equations for the gas interior coupled to diffusive transport, predicting sharp shock-like features near the minimum radius.
Hydrodynamic simulations of SL collapse, including the formation and propagation of compression waves inside the bubble, were carried out by Moss et al.~\cite{dabh8,dabh11} using the full Euler equations, and by Popinet and Zaleski~\cite{dabh12} for axisymmetric non-spherical bubble geometries.
A pedagogical treatment of the single-bubble dynamics in the continuum framework, including a systematic comparison of ODE models of increasing complexity, is given by Vignoli et al.~\cite{dabh13}.

Water vapor trapped inside the bubble during collapse plays a physically important role that purely noble gas models omit.
For instance, Storey and Szeri~\cite{storey2000water} and Toegel et al.~\cite{toegel2000does} demonstrated that the endothermic dissociation of H$_2$O provides a significant internal energy sink that limits the peak temperature and has a measurable effect on the SL emission spectrum.
Moss et al.~\cite{moss2000new} identified a new damping mechanism in strongly collapsing bubbles and proposed a modified RP equation that correctly predicts the rapid decay of post-collapse rebounds observed in SL experiments, resolving a longstanding deficiency of the classical RP equation, whose solutions persist until the next acoustic cycle.
Sensitivity studies by Vazquez and Putterman~\cite{vazquez2000temperature} and Yasui~\cite{yasui2001effect} quantified how the ambient liquid pressure and temperature, respectively, shift the collapse intensity and emission brightness, helping to anchor continuum models to measurable experimental observables.
More recent mesh-based simulations~\cite{lee2022numerical} of acoustic cavitation of bubbles in water have incorporated explicit water-vapor chemistry and spatial resolution of species and energy inside the bubble, further underscoring the importance of vapor dissociation in limiting peak temperatures and controlling radical production.

Continuum and hybrid radiative models have also been used to interrogate the ionization state and light emission mechanism of SL more directly.
Moss et al.~\cite{moss1999computed} computed optical emission from a partially ionized collapsing bubble including opacity effects; Hammer and Frommhold~\cite{hammer2002light} extended the analysis to rare-gas bubbles containing water vapor; and An~\cite{an2006mechanism} incorporated chemistry, ionization, and competing radiative channels into a unified single-bubble SL emission model.
These theoretical predictions are supported by direct experimental evidence: Flannigan and Suslick~\cite{flannigan2005plasma,flannigan2010inertially} and Flannigan et al.~\cite{flannigan2006measurement} established transient plasma formation in argon bubbles spectroscopically; Chen et al.~\cite{chen2008timeresolved} resolved the time-dependent emission structure; and Khalid et al.~\cite{khalid2012opacity} and Kappus et al.~\cite{kappus2013energy} characterized opacity effects and the energy budget of the light pulse, respectively.
At the interface level, molecular simulation studies of vapor--liquid kinetic boundary conditions have shown that accommodation-like departures from simple Maxwell reflection arise naturally at nonequilibrium liquid interfaces~\cite{ishiyama2005kinetic,ishiyama2013nonequilibrium}, underscoring the physical importance of the wall-coupling parameter $\alpha_t$ explored in Section~\ref{sec:results_boundary}.

% -----------------------------------------------------------------------
\subsection{Molecular Dynamics Simulation of Sonoluminescence}
% -----------------------------------------------------------------------

Molecular dynamics provides a complementary perspective on SL collapse: discrete particle trajectories naturally resolve shock formation, local ionization, and statistical fluctuations without requiring closure assumptions for the equation of state or transport coefficients.
The coupling between the MD gas interior and the surrounding liquid has been handled exclusively through a hybrid RP-MD (or KM-MD) approach, in which the MD simulation supplies the instantaneous gas pressure at the wall and the continuum equation advances the bubble radius.

The first MD study of SL was by Ruuth, Putterman, and Merriman~\cite{ruuth2002molecular}, who modeled the gas interior as a hard-sphere noble gas driven by a Rayleigh--Plesset piston.
Their event-based MD (EBMD) implementation, using fast tree-based collision-scheduling algorithms, reached up to $10^6$ ensemble particles and included a collision-energy threshold ionization model---the predecessor of the algorithm adopted in our work.
The study confirmed the formation of a central hot spot and estimated that the light emission duration scales with the ambient bubble radius.
The same work noted that between $10^9$ and $10^{10}$ physical particles would be required to simulate a realistic SL bubble without ensemble-particle scaling, a gap that remains largely open to this day.
Bass et al.~\cite{bass2008symmetry,bass2008mass} subsequently developed a symmetry-reduced hard-sphere MD framework for spherically symmetric bubble implosion, enabling simulations of xenon bubbles with up to $5\times10^7$ particles and revealing strong mass segregation during rapid collapse\footnote{Note, the actual particle counts in these simulations are on the order of $5.5 \times 10^4$--$1.5\times 10^6$; they simulate one cone of a bubble, which by symmetry they argue gives results equivalent to simulating a bubble with $5\times10^7$ particles, but since they only simulate a cone of the bubble their total particle counts are never that large.}.
These studies showed that algorithmic specialization can push SL bubble MD to much larger particle counts, but only by assuming exact spherical symmetry and remaining within the event-driven hard-sphere paradigm.

Kim, Kwak, and Kim~\cite{kim2007molecular} coupled a hard-sphere EBMD gas interior to the full Kwak--Na continuum model~\cite{kwak1997physical}, enabling a direct comparison between MD trajectories and PDE predictions for a xenon bubble in sulfuric acid at $10^6$ particles.
Their study found good qualitative and quantitative agreement between the MD and continuum results for the velocity, temperature, and pressure evolution, validating the hybrid KM-MD framework.
Among the most comprehensive of prior EBMD studies is Schanz et al.~\cite{dabh10}, who augmented the hard-sphere gas model with water vapor, vapor chemistry (dissociation, radical formation), and separate mass and energy transfer channels at the bubble wall.
Their implementation reached up to $10^7$ ensemble particles—though most production runs used $10^6$—and systematically explored the accommodation coefficient, driving amplitude, and noble-gas species, establishing a reference dataset that we use for direct comparison in Section~\ref{sec:results_compare}.
Ramsey~\cite{Ramsey2013EnergeticCavitation} examined energetic cavitation collapse, including sonoluminescence, from the perspective of scalable simulation and identified GPU parallelism as a key enabling technology for the next generation of large-scale SL computations.

Despite their respective contributions, prior SL MD studies share a set of structural limitations that motivate the present work.
\emph{Interaction model:}
Most prior large-scale SL MD simulations use hard spheres, and even the short-range model of Kim et al.~\cite{kim2007molecular} provides no natural route to including long-range Coulomb forces between the charged carriers produced by ionization.
At the extreme temperatures reached during SL collapse, the ionization fraction is non-negligible~\cite{ruuth2002molecular,dabh10}, and the Coulomb interaction between charge carriers exerts a physically meaningful restoring pressure that ought to be incorporated.
\emph{Scalability:}
EBMD requires maintaining a global priority queue of future collision events, an operation that is inherently sequential in time and difficult to parallelize across distributed-memory nodes or GPUs.
None of the prior EBMD implementations exploits GPU acceleration.
\emph{Particle count:}
The maximum ensemble size achieved in prior work ($10^7$ in Schanz et al.~\cite{dabh10}) falls two to three orders of magnitude short of particle counts seen in real-world experiments; while our present results ($N_\mathrm{ensem} = 10^8$, Section~\ref{sec:results_scale}) do not fully close this gap, we obtain an order-of-magnitude increase over state of the art---despite paying substantial additional costs for incorporating additional modeling terms.

The present work addresses each of these limitations by adopting a time-based MD (TBMD) formulation with a soft-sphere Lennard--Jones potential supplemented by a damped-shifted-force Coulomb interaction~\cite{fennell2006ewald}, full multi-stage ionization, and a KM-coupled continuum boundary, implemented in a scalable framework that reaches $10^8$ ensemble particles.
To our knowledge, this is the first application of TBMD to single-bubble sonoluminescence.

% -----------------------------------------------------------------------
\subsection{Large-Scale Molecular Dynamics Infrastructure}
% -----------------------------------------------------------------------

The computational demands of large-scale particle simulations have driven sustained development of general-purpose MD packages and parallelization strategies.
The LAMMPS package~\cite{thompson2022lammps,plimpton1995fast} provides a flexible, domain-decomposed framework for particle-based simulation that scales efficiently across thousands of CPU cores and has well-optimized GPU force kernels.
GPU acceleration for general MD was studied systematically by Glaser et al.~\cite{dabh17}, who demonstrated near-linear strong scaling across GPU clusters for systems of tens of millions of particles.
At the exascale frontier, the EXAALT project~\cite{dabh14} targets atomistic simulations across macroscopic length scales, ambitiously targeting billion-particle MD.
Large-scale MD has also been applied directly to bubble collapse: Vedadi et al.~\cite{vedadi2010shock} simulated shock-induced nanobubble collapse in water using reactive MD; Chen et al.~\cite{chen2023bubble} systematically studied system-size effects in bubble collapse MD with up to 4.5 billion water molecules; and Asano~\cite{asano2026cavitation} recently demonstrated 100-billion-atom acoustic cavitation MD on Fugaku, marking the current frontier of large-scale bubble simulation.
At the particle-method level, the use of simulated or ensemble particles to represent many physical particles is standard in particle-in-cell and DSMC methodologies~\cite{hockney2021computer,bird1994molecular}, and weighted-particle extensions have been used to control fluctuations for rare ionized species in weakly ionized flows~\cite{fang2020dsmc}.
Our ensemble-particle formulation is closest in spirit to this broader macro-particle tradition, although the specific scaling laws are tailored here to dense, collapsing LJ--Coulomb gas dynamics.
The present work is based on a version of LAMMPS with significant extensions for SBSL simulation, including ionization, wall coupling, and ensemble-particle scaling.
Our SL framework therefore inherits all the parallelization, scalability, and modular aspects of LAMMPS.
%we extend the reach of SL molecular simulation to a resolution regime previously inaccessible with event-based codes.
\section{Theory}
\label{sec:theory}

The dynamics of a sonoluminescing bubble couple the radial motion of the bubble wall, the energy exchange between the hot compressed gas and the surrounding liquid, and the spatiotemporal distribution of temperature and pressure inside the bubble.
We adopt the theoretical model developed by Kwak and collaborators~\cite{kwak1995aspect,kwak1997physical} as the continuum reference framework, which resolves each of these components through a coupled system of ordinary and partial differential equations.
The model serves two roles in this work: it provides the initial and boundary conditions used to initialize the molecular dynamics simulation (Section~\ref{sec:sim_proc}), and it supplies the continuum baseline against which MD results are compared throughout Section~\ref{sec:results}.
No single prior publication has collected a complete description of the numerical solution procedure for this model, so we present the governing equations and our implementation in the following subsections.

\subsection{Bubble motion}
The radial dynamics of the bubble wall are governed by the Keller--Miksis equation~\cite{keller1980bubble}, which accounts for liquid compressibility and acoustic radiation damping:
 \begin{equation}
\frac{d R_b}{dt}=U_b ,
\label{eq:PDE:Rb}
 \end{equation}
\begin{equation}
\left(1-\frac{U_b}{C}\right)R_b \frac{dU_b}{dt}
+\frac{3}{2}U_b^2\left(1-\frac{U_b}{3C}\right)
=\frac{1}{\rho_{\infty}}\left(1+\frac{U_b}{C}
+\frac{R_b}{C}\frac{d}{dt}\left[P_B-P_s\left(t+\frac{R_b}{C}\right)-P_{\infty}\right]\right) ,
\label{eq:PDE:ddRb}
\end{equation}
where $R_b$ is the bubble radius, $U_b$ is the bubble wall velocity, $C$ is the sound speed in the liquid, and $\rho_{\infty}$ is the ambient liquid density. The pressure due to the driving acoustic wave is $P_s(t)=-P_A \sin\omega t$, where $P_A$ is the driving pressure amplitude and $\omega=2\pi f$ with $f$ the driving frequency. The liquid pressure outside the bubble wall $P_B(t)$ is related by the pressure inside the bubble wall $P_b(t)$:
\begin{equation}
P_B=P_b(t)-\frac{2\sigma_s}{R_b}-\frac{4 \mu U_b}{R_b} ,
\label{eq:out_press}
\end{equation}
where $\sigma_s$ and $\mu$ are the surface tension and dynamic viscosity of the liquid medium.
\subsection{Liquid dynamics}
Although mass transfer through the bubble interface (evaporation and condensation of liquid molecules) is neglected, energy exchange between the gas and the surrounding liquid must be retained because it heats the liquid shell and thereby affects the predicted temperature at the bubble wall.
Heat transfer is assumed to occur through a thermal boundary liquid shell of thickness $\delta(t)$, governed by~\cite{ryu1992chaotic}:
\begin{equation}
\left[1+\frac{\delta}{R_b}+\frac{3}{10}\left(\frac{\delta}{R_b}\right)^2\right]
=\frac{6 \alpha}{\delta}
-\left[2\frac{\delta}{R_b}+\frac{1}{2}\left(\frac{\delta}{R_b}\right)^2\right]\frac{dR_b}{dt}
-\delta\left[1+\frac{1}{2}\frac{\delta}{R_b}+\frac{1}{10}\left(\frac{\delta}{R_b}\right)^2\right]\frac{1}{T_{bl}-T_{\infty}}\frac{dT_{bl}}{dt} ,
\label{eq:PDE:delta}
\end{equation}
where $\alpha$ is the thermal diffusivity of the liquid, $T_{bl}$ is the liquid temperature at the bubble wall, and $T_{\infty}$ is the liquid temperature far away from the bubble. 

\subsection{Temperature distribution in the liquid thermal boundary layer}
The temperature profile within the liquid boundary layer is assumed to be quadratic~\cite{theofanous1969theoretical}:
\begin{equation}
\frac{T-T_{\infty}}{T_{bl}-T_{\infty}}=(1-\xi)^2 ,
\end{equation}
where $\xi=(r-R_b)/\delta$ and $T_{bl}$ is again the temperature at the bubble wall. This quadratic profile satisfies the boundary conditions:
\begin{equation}
     T(R_b,t)=T_{bl}, \quad T(R_b+\delta,t)=T_{\infty}, \quad \left(\frac{\partial T}{\partial r}\right)_{r=R_b+\delta}=0 .
\label{eq:temperature boundary}
\end{equation} 
The temperature profile given in Eq.~\eqref{eq:temperature boundary} is valid provided the characteristic time scale of bubble evolution is much longer than the translational relaxation time of the molecules~\cite{byun2004bubble} yet comparable to or longer than the vibrational relaxation time~\cite{kwak1995bubble}.
\subsection{Temperature and pressure distribution inside the bubble}
The gas interior can be described in two ways: via a baseline uniform model, or via a nonuniform correction of said model.
We first solve for a spatially uniform baseline state, in which the density and pressure are taken to be uniform inside the bubble---an approximation that is reasonable except near the bubble's minimum radius.
Spatial nonuniformity can then be reconstructed as a secondary correction to this baseline solution, accounting for the nonuniform density and pressure induced by bubble wall acceleration, a transient effect that becomes significant on nanosecond time scales near collapse.

\paragraph{Uniform baseline}
The thermal conductivity of the gas inside the bubble is assumed to vary linearly with temperature:
\begin{equation}
k_g=AT+B .
\label{eq:conductivity}
\end{equation}
A linear fit to the conductivity data of Boulos et al.~\cite{boulos2013thermal} provides a good approximation in the range $200\,\mathrm{K} < T < 3000\,\mathrm{K}$~\cite{prosperetti1988nonlinear}. In practice, the linear form is applied up to $25{,}000\,\mathrm{K}$, above which $k_g$ is capped at $k_g' = 5\,\mathrm{W\,m^{-1}\,K^{-1}}$. Applying Fourier's law with this conductivity model yields the following radial temperature profile under the uniform assumption~\cite{kwak1995bubble}:
\begin{equation}
T_b(r)=\frac{B}{A}\left(-1+\sqrt{\left (1+\frac{A}{B}T_{b0}\right)^2-2\eta\left(\frac{A}{B}\right)(T_{bl}-T_{\infty})\left(\frac{r}{R_b}\right)^2 }\right) .
\end{equation}
The wall temperature $T_{bl}$ is obtained by applying the boundary conditions of Eq.~\eqref{eq:temperature boundary}:
\begin{equation}
T_{bl}=-\frac{B}{A}(1+\eta)+\left( \frac{B}{A}\right) \sqrt{(1+\eta)^2+2\left(\frac{A}{B}\right)\left(T_{b0}+\frac{A}{2B}T_{b0}^2+\eta T_{\infty}\right)} ,
\end{equation}
\begin{equation}
\eta=\frac{R_b}{\delta}\frac{k_1}{B} .
\label{eq:eta}
\end{equation}
Applying energy conservation to the uniformly distributed gas yields evolution equations for the center temperature and pressure:
\begin{equation}
\frac{dT_{b0}}{dt}=-\frac{3(\gamma-1)T_{b0}}{R_b}\frac{dR_b}{dt}-\frac{6(\gamma-1)k_1(T_{bl}-T_{\infty})T_{b0}}{\delta R_b P_{b0}} ,
\label{eq:PDE:Tb0}
 \end{equation}
\begin{equation}
\frac{dP_{b0}}{dt}=-\frac{3\gamma P_{b0}}{R_b}\frac{dR_b}{dt}-\frac{6(\gamma-1)k_1(T_{bl}-T_{\infty})}{\delta R_b} ,
\label{eq:PDE:Pb0}
\end{equation}
where $T_{b0}$ and $P_{b0}$ are the gas temperature and pressure at the bubble center, respectively, $\gamma$ is the specific heat ratio, and $k_1$ is the thermal conductivity of the liquid.

\paragraph{Nonuniform correction}
The abrupt temperature rise and subsequent rapid quenching driven by bubble wall acceleration occur on a time scale distinct from the conduction-dominated evolution captured by the uniform baseline.
This motivates incorporating a nonuniform correction.
Building on that baseline rather than treating the interior pressure as spatially constant, this correction decomposes the gas density into a spatially uniform part $\rho_0$ and a radially dependent correction $\rho_r$:
\begin{align}
\begin{split}
\rho_g &=\rho_0+\rho_r \\
u_g &=\frac{\dot{R_b}}{R_b}r \\
P_b&=P_{b0}(t)-\frac{1}{2}(\rho_0+\frac{1}{2}\rho_r)\frac{\ddot{R_b}}{R_b}r^2 ,
\label{eq:density}
\end{split}
\end{align}
where $\rho_0 R_b^3=\mathrm{const}$, $\rho_r=ar^2/R_b^5$, and a dot denotes a time derivative. The constant $a$ is related to the total gas mass $m$ inside the bubble by
\begin{equation}
\frac{a}{m}=\frac{5}{4\pi}(1-N_{BC}) ,
\end{equation}
where $N_{BC}=(P_{b0}R_b^3/T_{b0})/(P_{\infty}R_0^3/T_{\infty})$ is a constant obtained from the uniform baseline and $R_0$ is the equilibrium radius. The nonuniform density and pressure distributions yield an inertial correction to the temperature profile:
\begin{align}
\begin{split}
T(r)&=T_b(r)+T'_b(r) \\
T'_b(r)&=-\frac{1}{40(\gamma-1)k_g'}\left(\rho_0+\frac{5}{21}\rho_r \right)\left[(3\gamma-2)\frac{\dot{R_b}\ddot{R_b}}{R_b^2}+\frac{\dddot{R_b}}{R_b}\right]r^4+C(t) \\
C(t)&=\frac{1}{20(\gamma-1)}[(3 \gamma-2)\dot{R_b}\ddot{R_b}R_b+\dddot{R_b}R_b^2]\times \left[\frac{\delta'}{k_l}\left(\rho_0+\frac{5}{14}\rho(R_b)\right)+\frac{R_b}{2 k_g'}\left(\rho_0+\frac{5}{21}\rho(R_b)\right)\right] .
\end{split}
\label{eq:nonuniform}
\end{align}
% The coefficient $C(t)$ is determined from a boundary condition $k_g' dT_b/dr=k_l dT_l/dr$ at the wall where $T_l$ is the temperature distribution in the thermal boundary layer with different thickness $\delta'$. 
% In the original non-uniform model of Kwak \& Na \cite{kwak1997physical}, the gas-side thermal layer thickness $\delta'$ is treated as a semi-empirical parameter, chosen (around 0.1 µm) to reproduce realistic post-collapse rebound in a purely continuum setting. 
% In our hybrid MD–continuum framework, radial heat transport inside the gas is resolved explicitly by MD, and the only remaining role of $\delta'$ is to characterize the thickness of the near-wall gas region that exchanges energy with the liquid.
% We therefore reinterpret $\delta'$ as an effective coupling-layer thickness and set
% $\delta'$ equal to the cutoff radius of the gas–wall interaction potential in physical units (Introduce in the Section \ref{sec:md}). 
% We find that this choice yields significantly better agreement with both the continuum reference solution and the MD observables than using  $\delta'=0.1 µm$, which overpredicts the collapse temperature in our setup.
The coefficient $C(t)$ is determined from the interfacial heat--flux
condition $k_g' \,\partial_r T_b = k_l \,\partial_r T_l$ at the bubble wall,
where $T_l$ denotes the temperature inside the quadratic liquid thermal
boundary layer of thickness $\delta'$. In the continuum nonuniform model
of Kwak \& Na~\cite{kwak1996hydrodynamic}, $\delta'$ is treated as a
semi--empirical gas--side thermal layer thickness, and
Eqs.~\eqref{eq:nonuniform} yield a perturbative inertial correction
$T_b'(r,t)$ that is intended to remain small relative to the
conduction--controlled background temperature $T_b(r,t)$.

However, when evaluated in the strongly driven SL regime considered here,
the authors found that
$T_b'(r,t)$ becomes comparable to or larger than $T_b(r,t)$ and may even
render the predicted gas temperature negative near the interface at
certain collapse stages, indicating that the linearized correction lies
outside its validity range. 
For this reason, we retain $T_b(r,t)$ as the
quantitative theoretical temperature profile used for comparison with MD
results, and regard further investigation into nonuniform correction models as future work.
%and regard $T_b'(r,t)$ only qualitatively as an indication of
%additional inertial nonuniformity beyond the regime of applicability of
%the perturbative continuum model.

\begin{figure}[htbp]
  \centering
  \subfloat[]
  {\includegraphics[width=0.48\textwidth]{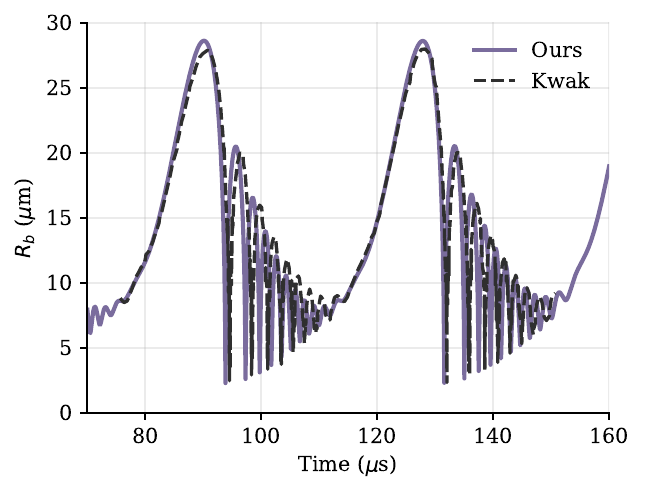}\label{fig:validation}}
  \subfloat[]
   {\includegraphics[width=0.48\textwidth]{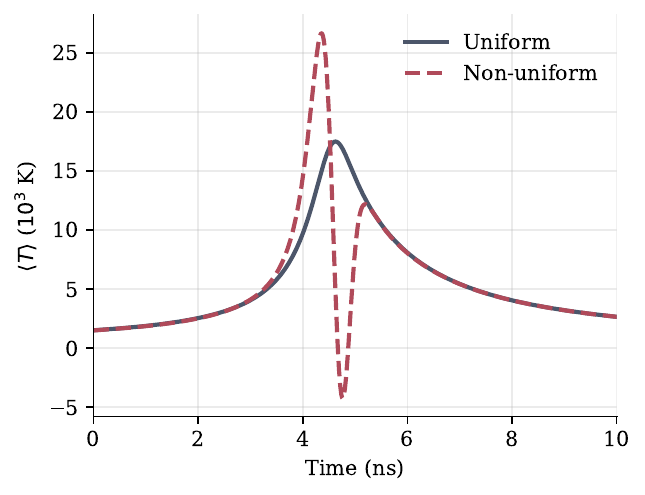}\label{fig:nonuniform_breakdown}}
 \caption{Validation and analysis of the Kwak continuum model.
 (a)~Bubble radius as a function of time computed with the present numerical solver (solid) compared against the theoretical radius--time curve of Kwak and Yang~\cite{kwak1995aspect} (dashed; hand-digitized from their Fig.~2) for an air bubble of equilibrium radius $R_0 = 8.5\,\mu$m driven at $P_A = 1.075\,$atm and $f = 26.5\,$kHz in water, the sub-threshold case reported in that reference. The drive-cycle period, collapse timing, and maximum radius are reproduced.
 (b)~Spatially averaged gas temperature $\langle T \rangle$ (in units of $10^3\,\mathrm{K}$) predicted by the uniform baseline and the nonuniform correction for the strongly driven argon--water SL configuration used in this work. Near collapse, the inertial correction $T_b'(r,t)$ drives the nonuniform prediction below zero, indicating that the perturbative expansion has left its regime of validity. For this reason, the uniform-baseline profile $T_b(r,t)$ is used as the quantitative continuum reference throughout the remainder of the paper.}
 \label{fig:uniform_non_uniform}
\end{figure}

We note that we verified our implementation against the continuum model of Kwak \& Na by reproducing their air--bubble predictions under the same acoustic and thermophysical conditions reported in Kwak \& Yang~\cite{kwak1995aspect}, obtaining excellent agreement for both the bubble radius evolution and the associated temperature field (Fig.~\ref{fig:validation}). However, when the same nonuniform formulation, including the inertial correction $T_b'(r,t)$ of Eqs.~\eqref{eq:nonuniform}, is applied to the strongly driven SL configuration considered in this work, the model produces unphysical results: at certain collapse stages it predicts negative gas temperatures near the bubble wall and substantially distorted thermal profiles (Fig.~\ref{fig:nonuniform_breakdown}). We conjecture that these pathologies arise because the quantity driving the correction term, $(3\gamma-2)\dot R_b\ddot R_b/R_b^2 + \dddot R_b/R_b$, becomes sufficiently large that $T_b'(r,t)$ is no longer a small perturbation to the conduction-controlled background temperature $T_b(r,t)$, violating the formal assumptions of the perturbative derivation in Kwak \& Na~\cite{kwak1996hydrodynamic}. For this reason, in our simulations, we retain the full continuum dynamics for the mean bubble evolution while restricting ourselves to using the uniform model  $T_b(r,t)$ as the quantitative theoretical temperature profile. This choice captures the dominant nonuniform conduction structure and avoids introducing the non-physical temperature artifacts that arise when $T_b'(r,t)$ is applied outside its regime of validity.

\subsection{Numerical solver}
\label{sec:numerical_solver}
Our numerical solver integrates the five-variable uniform baseline system (Eqs.~\eqref{eq:PDE:Rb}--\eqref{eq:PDE:Pb0}) using MATLAB's \texttt{ode45} solver.
To ensure consistency with our molecular dynamics experiments (detailed in subsequent sections), we initialize the system with the following conditions: $R_b = R_0 = 4.5\,\mu$m, $U_b = 0$ m/s, $P_b = P_{\infty} = 101325$ Pa, $T_{bl} = T_{b0} = T_{\infty} = 300.0$ K, and $\delta = 0.3 R_0$. The remaining simulation parameters are set as follows: $f = 26.5$ kHz, $\gamma = \frac{5}{3}$, $P_A = 1.3 P_{\infty}$, $C = 1481$ m/s, $\mu = 0.001$ Pa$\cdot$s, $\sigma_s = 0.072$ N/m, $\rho_{Ar} = 1.603$ kg/m$^3$, $A = 2.682 \times 10^{-5}$ W/mK$^2$, $B = 1.346 \times 10^{-2}$ W/mK, $k_1 = 0.61$ W/mK, and $N_{BC} = 1.316$.
The nonuniform inertial correction $T_b'(r,t)$ of Eqs.~\eqref{eq:density} and \eqref{eq:nonuniform} is evaluated post hoc from the baseline solution solely to generate Fig.~\ref{fig:nonuniform_breakdown}; as discussed above, it is not used as a quantitative prediction.

To maintain numerical stability and accuracy, we employ a two-stage adaptive time-stepping scheme.
First, we perform a preliminary simulation with a fixed timestep of $10^{-10}$\,s to obtain a coarse estimate of the time at which the bubble reaches its minimum radius.
Based on this estimate, we then rerun a refined simulation with a timestep of $10^{-15}$\,s, starting approximately 5\,ns before the coarse minimum-radius time and running for a total of 10\,ns.
Throughout this work, $t_{\min}$ denotes the minimum-radius time identified by this fine-run simulation.
Because the timestep decreases by five orders of magnitude between the two stages, the refined $t_{\min}$ does not in general coincide with the midpoint of the stage-2 window; in our configuration it falls at approximately 4.6\,ns from the start of that window rather than at the expected 5\,ns.
This offset is reflected in the simulation time axes of the results section.

\section{Molecular Dynamics Simulations}
\label{sec:md}
This section describes the molecular dynamics component of our hybrid continuum--MD framework. The principal challenge in simulating sonoluminescence is the need to track large particle ensembles through extreme thermodynamic states on nanosecond time scales. We address this using the LAMMPS molecular dynamics package~\cite{thompson2022lammps} as the computational backbone, extended with custom modules for ionization, gas--wall energy exchange, and Keller--Miksis wall coupling.
%To be specific, finding the balance between model complexity and the time consumption.
\subsection{Hard-sphere and soft-sphere models}
In the hard-sphere model, each atom is represented as a rigid sphere that interacts only through instantaneous, perfectly elastic binary collisions; no interaction force acts between particles at any other time. The soft-sphere model, by contrast, assigns each particle a continuous, spherically symmetric potential (e.g., Lennard--Jones) with a finite cutoff. Although the hard-sphere model reproduces noble gas viscosity accurately~\cite{bird1994molecular} and shows little difference from soft-sphere results in dilute conditions~\cite{elghannay2019evaluation}, its validity in a dense, strongly compressed system near the minimum bubble radius is unclear. Furthermore, a continuous potential provides a natural route to include long-range Coulomb forces between charged carriers produced during ionization, enabling a more physically complete simulation.
%Hard sphere & event base \cite{smallenburg2022efficient}
% Soft sphere & event base \cite{muller2013event}

\subsection{Time- and event-based molecular dynamics}
Time-Based Molecular Dynamics (TBMD) relies on integrating the equations of motion using discrete time steps, such as through the Verlet or leapfrog algorithms. At each time step, the forces on the particles are computed, and their positions and velocities are updated accordingly. Event-Based Molecular Dynamics (EBMD), on the other hand, takes a different approach by focusing only on events of interest, such as collisions between particles or changes in potential energy states. Instead of updating particle positions at every small time step, the algorithm predicts the time of the next event and skips directly to it. In the SL context, the timestep and initial conditions must be chosen carefully to achieve a stable and accurate simulation.
There is a one-to-one correspondence with the preceding subsection: EBMD is based on the hard-sphere model, whereas TBMD is based on the soft-sphere model.
The remainder of this section assumes TBMD and soft spheres for the particles.

We note that in the context of SBSL, a potentially drastic downside of the EBMD approach is that as millions or even billions of particles are compressed into a microscopic bubble, the number of collisions increases exponentially, incurring great computational cost to perform searches for the next collision.
In this highly compressed regime, TBMD may also have a drawback in that two or more particles may non-physically pass through each other if the time step is too large (i.e., collisions are missed).
However, we take the view that a small number of missed collisions is an acceptable price to pay for being able to take a relatively large time step, and our numerical results demonstrate that our TBMD approach produces results that align with the literature while enabling substantially larger particle counts.

\subsection{Gas dynamics inside the bubble}
\label{sec:gas_dynamics}
In contrast to the hard-sphere model, where collisions are resolved only at discrete events based on energy and momentum conservation, the present approach requires computing the pairwise interaction potential at each timestep. The total pair interaction energy combines a short-range Lennard--Jones term and a Coulombic contribution computed via the damped shifted force (DSF) method~\cite{fennell2006ewald}:
\begin{equation}
E_{\text{pair}}(r) = E_{\text{LJ}}(r) + E_{\text{Coul-DSF}}(r),
\end{equation}
where each contribution is computed with its own cutoff radius.

\paragraph{Lennard--Jones potential}
Short-range molecular interactions are governed by the Lennard--Jones 12-6 potential, which accounts for van der Waals forces:  

\begin{equation}  
E_{\text{LJ}}(r) =  
\begin{cases}  
4 \epsilon \left[\left(\frac{\sigma}{r}\right)^{12}-\left(\frac{\sigma}{r}\right)^{6} \right], & r < r_{\text{LJ}} \\  
0, & r \geq r_{\text{LJ}}  
\end{cases}  
\end{equation}  
where \( r \) is the interatomic separation, \( \epsilon \) is the depth of the potential well, and \( \sigma \) is the characteristic length scale at which the potential energy is zero. The interactions are truncated at a cutoff distance \( r_{\text{LJ}} \), beyond which the potential is set to zero. For argon we use $\sigma = 3.401\,\text{\AA}$ and $\varepsilon/k_B = 116.81\,\text{K}$; all force-field and cutoff parameters are collected in Table~\ref{tab:params}.  

\paragraph{Coulombic interaction with the DSF method}
For charged particles, the electrostatic interaction is modeled using the damped shifted force (DSF) potential~\cite{fennell2006ewald}, which provides a computationally efficient alternative to long-range summation methods such as Ewald~\cite{ewald1921berechnung} or particle--particle particle--mesh (PPPM)~\cite{hockney2021computer} while ensuring smooth truncation and local charge neutrality within a cutoff radius $r_{\text{Coul}}$:

\begin{equation}  
E_{\text{Coul-DSF}}(r) =  
\begin{cases}  
q_i q_j \left[ \frac{\operatorname{erfc}(\alpha r)}{r} - \frac{\operatorname{erfc}(\alpha r_{\text{Coul}})}{r_{\text{Coul}}} + \left(\frac{\operatorname{erfc}(\alpha r_{\text{Coul}})}{r_{\text{Coul}}^2} + \frac{2\alpha}{\sqrt{\pi}} \frac{e^{-\alpha^2 r_{\text{Coul}}^2}}{r_{\text{Coul}}} \right)(r - r_{\text{Coul}}) \right], & r < r_{\text{Coul}} \\  
0, & r \geq r_{\text{Coul}}  
\end{cases}  ,
\end{equation}  
where $q_i$ and $q_j$ are the particle charges, $\alpha$ is the damping parameter, and $\operatorname{erfc}(\cdot)$ is the complementary error function. We set the damping parameter to $\alpha = 3/r_{\text{Coul}}$, which places the Gaussian screening width at one-third of the cutoff so that the residual interaction at the cutoff, $\operatorname{erfc}(\alpha r_{\text{Coul}}) = \operatorname{erfc}(3) \approx 2.2\times10^{-5}$, is a negligible fraction of the bare Coulomb value. In the production runs, both cutoffs are set equal: $r_{\text{Coul}} = r_{\text{LJ}} = 2\,d_{\text{ensem}} \approx 2.15\,\sigma_{\text{ensem}}$, ensuring Coulomb interactions are resolved across the full LJ range. The ionization reactive cutoff is $r_{\text{ion}} = g^{1/2}d$, where $d = 3.66\,\text{\AA}$ is the argon atomic diameter and $g = N_{\text{real}}/N_{\text{ensem}}$.

%\cite{donev2005neighbor} % a good refernce for event-based introdution
\subsection{Gas ionization inside the bubble}
Near the minimum bubble radius, the extreme pressure and temperature are sufficient to ionize a significant fraction of the gas. Ionization removes kinetic energy from the gas and thereby limits the peak temperature, making its proper treatment essential for accurate predictions. We follow a strategy analogous to that of Ruuth et al.~\cite{ruuth2002molecular}: whenever the combined kinetic energy $E = E_i + E_j$ of a candidate pair exceeds the ionization potential $\chi$ of the less-ionized atom, ionization is assumed to occur with certainty. Both particles' velocities are rescaled by the same factor $s = \sqrt{2(E-\chi)/(3E)}$, giving each particle $k \in \{i,j\}$ a post-ionization kinetic energy
\begin{equation}
E_{k,\mathrm{new}} = \frac{2(E - \chi)}{3E}\,E_k ,
\label{eq:ionize}
\end{equation}
where $E_k$ is the pre-event kinetic energy of particle $k$. The total energy extracted from the pair is $\chi + \tfrac{1}{3}(E-\chi)$: the first term is the ionization potential, and the second accounts for the freed electron taking one-third of the remaining kinetic energy as it thermalizes with the gas---the two-loss mechanism identified by Ruuth et al.~\cite{ruuth2002molecular}. The model tracks how many electrons each particle loses, enabling the use of successive ionization energies for future ionization events. This also helps determine local ionization levels and properly account for Coulomb forces. A check for potential ionization events is conducted at every timestep. Specifically, if two atoms, \( i \) and \( j \), are within the ionization reactive cutoff $r_\mathrm{ion}$ of each other and their combined kinetic energy exceeds the ionization energy of at least one of the atoms, the pair becomes a candidate for ionization. The reactive cutoff $r_\mathrm{ion}$ is an independent parameter, distinct from the Lennard--Jones and Coulomb cutoffs, that sets the range within which a collision can trigger ionization. During the compression phase of sonoluminescence, each atom may encounter multiple neighboring atoms eligible for ionization. To resolve this, each atom examines its list of potential ionization partners and selects the one with the shortest distance as its ``sole'' ionization candidate. 

Once this labeling is complete, a pair \( (i, j) \) is deemed an ``eligible'' ionization pair if and only if \( i \) labels \( j \) as its sole ionization partner and vice versa. If the combined kinetic energy is sufficient to ionize either \( i \) or \( j \), the model prioritizes ionizing the atom with the lower ionization level, thereby favoring the atom that requires less energy to ionize further. This approach does not aim to strictly replicate the continuous nature of ionization in the physical world; instead, it provides a discrete approximation, with finer timesteps yielding more accurate representations of ionization dynamics.

\subsection{Gas-wall dynamics and energy transfer}

In modeling liquid--gas interactions, the surrounding liquid is treated as a continuum, and gas--liquid coupling arises from collisions of gas molecules with the bubble wall. The gas--liquid interface is represented by a repulsive Lennard--Jones wall potential of the same 12-6 form described in Section~\ref{sec:gas_dynamics}. Because the liquid is treated as a continuum, heat exchange between gas and liquid occurs at the boundary through molecular--wall collisions. Each collision leads to an adjustment of particle velocities, thereby capturing thermal effects at the boundary.

The heat energy \(P_Q\) transferred by \(\Delta N\) particles colliding with the wall within a timestep \(\Delta t\) is given by:
\begin{equation}
P_Q = \frac{3}{2} k_B \frac{\Delta N}{\Delta t} (T_i - T_r)
     = \frac{3}{2} k_B \alpha_t \frac{\Delta N}{\Delta t} (T_i - T_w),
\label{eq:heatflux}
\end{equation}
where \(T_i\) is the temperature of the incoming particles, \(T_w\) is the wall temperature, and \(k_B\) is the Boltzmann constant. The thermal accommodation coefficient \(\alpha_t\) governs the degree of heat conduction between the gas and liquid. The temperature of particles reflected from the wall, \(T_r\), is then
\begin{equation}
T_r = (1 - \alpha_t)\,T_i + \alpha_t\,T_w.
\end{equation}

By selecting an appropriate \(\alpha_t\), one can control the thermal boundary condition: \(\alpha_t = 1\) corresponds to an isothermal boundary, whereas \(\alpha_t = 0\) corresponds to a completely adiabatic boundary. Because the wall potential is continuous, a single physical collision spans multiple timesteps. To avoid applying the thermal energy transfer multiple times within one contact, we track for each molecule $p$ the timestep $t^p$ of its most recent wall contact, updated whenever the molecule lies within the gas--wall interaction cutoff. Energy transfer is applied at the onset of each new contact event, identified as the first timestep for which $t^p$ advances by more than one step from its previous value.

\subsection{Energy redistribution within the liquid shell}
The total energy stored in the liquid shell is updated at each timestep to account for three contributions:
\begin{equation}
E_{\mathrm{new}} = E_{\mathrm{old}} + P_Q\,\Delta t + E_v + E_{\mathrm{acoustic}},
\label{eq:shell_energy}
\end{equation}
where $P_Q$ is the gas-to-liquid heat flux at the bubble wall (Eq.~\eqref{eq:heatflux}), $E_v$ captures the energy change due to advection as the liquid shell expands or contracts, and $E_{\mathrm{acoustic}}$ is the work performed by the external acoustic field on the shell.

The advection term is
\begin{equation}
E_v = C_l \rho_\infty T_\infty (V_{\mathrm{new}} - V_{\mathrm{old}}),
\label{eq:Ev}
\end{equation}
where $C_l$ is the specific heat capacity of the liquid, $\rho_\infty$ is the ambient liquid density, $T_\infty$ is the ambient temperature, and $V_{\mathrm{new}} - V_{\mathrm{old}}$ is the change in volume of the liquid shell. The acoustic work term is
\begin{equation}
E_{\mathrm{acoustic}} = P_s(t)\cdot 4\pi R_b^2\,\dot{R}_b\,\Delta t,
\label{eq:acoustic_work}
\end{equation}
where $P_s(t) = -P_A\sin\omega t$ is the acoustic driving pressure. This term ensures that the energy bookkeeping is thermodynamically consistent across the full acoustic cycle, regardless of the duration of the MD simulation window.

Given the updated total energy in the liquid shell, the bubble wall temperature $T_{bl}$ is recovered through the quadratic temperature distribution of Eq.~\eqref{eq:temperature boundary}.
At each timestep, the MD-computed gas pressure at the bubble wall $P_b$ and the heat flux $P_Q$ are returned to the continuum solver to update $R_b$, $U_b$, and $T_{bl}$, closing the hybrid loop.

\subsection{End-to-end algorithm for one timestep}
\label{sec:md_algorithm}

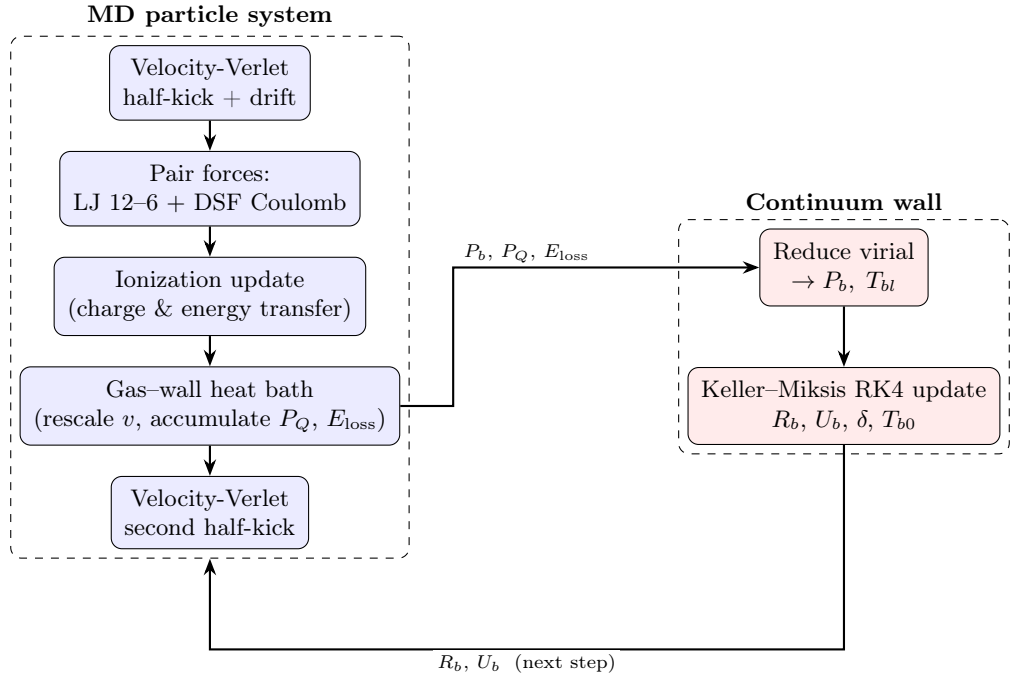
\begin{figure}[htbp]
  \centering
  \begin{tikzpicture}[
      font=\small,
      box/.style={draw, rounded corners, align=center, inner sep=5pt, minimum height=8mm},
      md/.style={box, fill=blue!8},
      cont/.style={box, fill=red!8},
      io/.style={box, fill=black!5},
      arr/.style={-{Stealth[length=2.2mm]}, thick},
      lab/.style={font=\scriptsize, fill=white, inner sep=1.5pt},
      every node/.style={align=center}
    ]
    % MD column
    \node[md] (verlet) {Velocity-Verlet\\half-kick + drift};
    \node[md, below=4mm of verlet] (pair) {Pair forces:\\LJ 12--6 $+$ DSF Coulomb};
    \node[md, below=4mm of pair] (ion) {Ionization update\\(charge \& energy transfer)};
    \node[md, below=4mm of ion] (wall) {Gas--wall heat bath\\(rescale $v$, accumulate $P_Q$, $E_\mathrm{loss}$)};
    \node[md, below=4mm of wall] (kick2) {Velocity-Verlet\\second half-kick};

    % Continuum block (pushed further right to clear the coupling label)
    \node[cont, right=38mm of wall] (km) {Keller--Miksis RK4 update\\$R_b,\,U_b,\,\delta,\,T_{b0}$};
    \node[cont, above=8mm of km] (press) {Reduce virial\\$\rightarrow P_b,\;T_{bl}$};

    % grouping rectangles
    \begin{scope}[on background layer]
      \node[draw, dashed, rounded corners, fit=(verlet)(pair)(ion)(wall)(kick2), label=above:{\textbf{MD particle system}}] (mdgroup) {};
      \node[draw, dashed, rounded corners, fit=(press)(km), label=above:{\textbf{Continuum wall}}] (cgroup) {};
    \end{scope}

    % internal MD arrows
    \draw[arr] (verlet) -- (pair);
    \draw[arr] (pair) -- (ion);
    \draw[arr] (ion) -- (wall);
    \draw[arr] (wall) -- (kick2);

    % continuum internal arrow: Reduce virial -> KM update (single downward arrow)
    \draw[arr] (press) -- (km);

    % MD -> continuum: go UP first, then RIGHT into the "Reduce virial" box,
    % so the line never crosses the KM box. Label on the vertical run.
    \draw[arr] (wall.east) -- ($(wall.east)+(7mm,0)$)
      |- (press.west)
      node[lab, pos=0.62, above] {$P_b,\,P_Q,\,E_\mathrm{loss}$};

    % continuum -> MD (next step): drop well below everything, ample clearance
    \coordinate (rety) at ($(mdgroup.south)+(0,-12mm)$);
    \draw[arr] (km.south) |- (rety -| km)
      -- (rety -| mdgroup.south)
      node[lab, midway, below] {$R_b,\,U_b$ \;(next step)}
      -- (mdgroup.south);
  \end{tikzpicture}
  \caption{Data flow over one timestep of the hybrid TBMD--continuum framework. The MD particle system (left) advances the gas atoms and measures the wall gas pressure $P_b$, heat flux $P_Q$, and collisional energy loss $E_\mathrm{loss}$; these drive the Keller--Miksis continuum update (right), which returns the new bubble radius $R_b$ and wall velocity $U_b$ that define the moving boundary for the next MD step. The coupling is staggered by one step.}
  \label{fig:coupling_schematic}
\end{figure}

The components described above---velocity-Verlet integration, the Lennard--Jones/DSF pair interaction, the discrete ionization model, the gas--wall heat bath, and the Keller--Miksis continuum solver---are advanced together once per MD timestep within a single LAMMPS run loop.
Figure~\ref{fig:coupling_schematic} shows how the data flows between the MD particle system and the continuum bubble-wall state, and Algorithm~\ref{alg:timestep} gives the complete step-by-step procedure for one timestep.
The two solvers are coupled in a staggered manner: the gas pressure $P_b$ and wall temperature $T_{bl}$ measured by the MD system at the current step are consumed by the continuum update, whose new wall radius $R_b$ and velocity $U_b$ define the moving boundary seen by the MD particles at the next step.

\begin{algorithm}[t]
\caption{One full timestep of the hybrid TBMD--continuum SL solver}
\label{alg:timestep}
\small
\setlength{\baselineskip}{0.92\baselineskip}
\begin{algorithmic}[1]
\Require particle states $\{\mathbf{x}_i,\mathbf{v}_i,q_i\}$; wall state $(R_b,U_b,\delta,T_{b0},T_{bl})$; pressures $P_b$, $P_b^{\text{old}}$ from previous step; timestep $\Delta t$
\Statex \textbf{// 1.\ Integrate particle motion (velocity Verlet, first half)}
\State $\mathbf{v}_i \gets \mathbf{v}_i + \tfrac{1}{2}\Delta t\,\mathbf{f}_i/m_i$;\quad $\mathbf{x}_i \gets \mathbf{x}_i + \Delta t\,\mathbf{v}_i$ \Comment{for all atoms $i$}
\State rebuild neighbor lists and exchange ghost atoms
\Statex \textbf{// 2.\ Ionization (discrete, before force accumulation)}
\For{each mutually labeled nearest pair $(i,j)$ with $r_{ij}<r_{\mathrm{ion}}$}
  \If{combined kinetic energy $\geq$ next ionization potential $\chi$}
    \State ionize lower-charged atom: increment $q$, subtract $\chi$ from the pair (Eq.~\eqref{eq:ionize}\,analogue)
  \EndIf
\EndFor
\State clear stale labels; relabel each atom's sole nearest ionization candidate for the next step
\Statex \textbf{// 3.\ Pairwise forces}
\For{each pair $(i,j)$ within cutoff}
  \State accumulate LJ 12--6 force (if $r_{ij}<r_{\text{LJ}}$) and DSF Coulomb force (if $r_{ij}<r_{\text{Coul}}$, with $\alpha=3/r_{\text{Coul}}$)
\EndFor
\Statex \textbf{// 4.\ Gas--wall coupling (\texttt{post\_force})}
\For{each atom $i$ within $r_c$ of the bubble wall at radius $R_b$}
  \State apply repulsive Lennard--Jones 12-6 wall force
  \If{$i$ is in first contact this collision}
    \State rescale $\mathbf{v}_i$ toward $T_{bl}$ by $\sqrt{1-\alpha_t+\alpha_t\,T_{bl}/T_i}$; \, accumulate $E_\mathrm{loss}$
  \EndIf
  \State accumulate pair/kinetic virial for the wall shell
\EndFor
\State MPI-sum the virial accumulators\footnotemark\ across all ranks $\Rightarrow$ gas pressure $P_b$ and gas-shell temperature
\State update liquid-shell energy ($E_\mathrm{loss}+$ advection $+$ acoustic work) $\Rightarrow$ wall temperature $T_{bl}$
\Statex \textbf{// 5.\ Integrate particle motion (velocity Verlet, second half)}
\State $\mathbf{v}_i \gets \mathbf{v}_i + \tfrac{1}{2}\Delta t\,\mathbf{f}_i/m_i$
\Statex \textbf{// 6.\ Continuum bubble-wall update (\texttt{shape\_update})}
\State advance the parallel continuum (theory) RK4\footnotemark\ of the Kwak model to obtain the liquid-shell thickness $\delta$
\State form $P_B = P_b - 2\sigma_s/R_b - 4\mu U_b/R_b$ from the measured $P_b$
\State RK4-advance Keller--Miksis Eqs.~\eqref{eq:PDE:Rb}--\eqref{eq:PDE:Pb0} by $\Delta t$ using $(P_B,P_B^{\text{old}})$ to obtain new $(R_b,U_b,T_{b0})$
\State $P_b^{\text{old}} \gets P_b$ \Comment{stage handoff for next step}
\Statex \textbf{// 7.\ Time-step control and output}
\State adapt $\Delta t$ within $[1,\,8]\,\mathrm{fs}$ (maximum displacement $\approx 0.07\,\text{\AA}$ per step), check halt condition, write thermo/dump output
\end{algorithmic}
\end{algorithm}
\footnotetext[3]{Per-rank partial sums of $\mathbf{r}_{ij}\cdot\mathbf{f}_{ij}$; their global sum gives the pressure via the virial theorem.}
\footnotetext[4]{The LAMMPS-coupled solver uses fixed-step RK4 because LAMMPS enforces a fixed MD timestep $\Delta t$ (see Fig.~\ref{fig:coupling_schematic}); the standalone continuum solver of Section~\ref{sec:numerical_solver} uses MATLAB's adaptive \texttt{ode45} instead.}
\section{Ensemble-Particle Representation and Scaling}
\label{sec:ep_scaling}

Even with our TBMD formulation, simulating the $N_\mathrm{real}\sim 10^{10}$ gas atoms inside a sonoluminescing bubble at fully atomistic resolution is computationally challenging.
We therefore represent the gas by $N_\mathrm{ensem} \ll N_\mathrm{real}$ \emph{ensemble particles} (EPs), each standing in for $g = N_\mathrm{real}/N_\mathrm{ensem}$ microscopic atoms that occupy similar positions and velocities in phase space.
The goal of the mapping is to preserve the dominant mesoscopic observables---temperature, short-range collisional hardness, pressure, and ionization statistics---while reducing the particle count to a more computationally tractable level.

Our proposed design principle is as follows.
Replacing $N_\mathrm{real}$ atoms with $N_\mathrm{ensem} = N_\mathrm{real}/g$ EPs reduces the number density by a factor of $g$.
To maintain the same local packing and interaction frequency, all characteristic length scales are enlarged by $g^{1/3}$, restoring the volume per particle to its physical value.
All energies are scaled by $g$ so that the energy per microscopic atom, and hence the temperature, is unchanged.
The following subsections apply this principle to each interaction in the model.

\subsection{Mass and kinetic energy}

Each EP aggregates the mass of $g$ atoms while leaving the velocity unchanged:
\begin{equation}
    m' = g\,m, \qquad v' = v.
    \label{eq:ep_mass}
\end{equation}
The per-EP kinetic energy is therefore $K' = \tfrac{1}{2}m'v^2 = g\,(\tfrac{1}{2}mv^2)$; since $v' = v$ and the temperature formula uses the microscopic mass $m$, the temperature is unchanged.
The Maxwellian velocity distribution and all characteristic thermal velocities and flight times are preserved exactly.

\subsection{Interaction potential scaling}

\paragraph{Lennard--Jones potential}
Reducing the number density by $1/g$ would soften short-range encounters and alter the virial pressure contribution.
To maintain the effective collisional hardness, the LJ parameters are rescaled as
\begin{equation}
    \sigma' = g^{1/3}\sigma, \qquad \epsilon' = g\,\epsilon, \qquad r_c' = g^{1/3} r_c.
    \label{eq:ep_lj}
\end{equation}
This enlarges the interaction range to the EP length scale and increases the well depth so that EP--EP collisions reproduce the local stress contribution of $g$ microscopic pairs.

\paragraph{Coulombic DSF interaction}
The length and damping parameter of the DSF potential follow the same geometric scaling,
\begin{equation}
    r_c' = g^{1/3} r_c, \qquad \alpha' = g^{-1/3}\alpha,
    \label{eq:ep_dsf_cutoff}
\end{equation}
which preserves the shape of the screened Coulomb field at the EP scale.
Charges are left unscaled ($q_i' = q_i$). Each EP has mass $m' = gm$ but carries a single ionic charge $q$, so without scaling its Coulomb acceleration would be $1/g$ that of the physical ion. Setting
\begin{equation}
    s_{\mathrm{pair}} = g,
    \label{eq:ep_dsf_scale}
\end{equation}
restores the correct ionic acceleration. This leaves the Coulomb virial undercounted by $g^{1/3}$; since the ionized fraction reaches at most 6--12\%, the impact on the total pressure is small.

\subsection{Ionization scaling}

Ionization encounters are screened by the reduced EP number density.
The reactive radius is enlarged as
\begin{equation}
    r_{\mathrm{ion}}' = g^{1/2} r_{\mathrm{ion}},
    \label{eq:ep_rion}
\end{equation}
which compensates for the lower encounter rate and yields an approximately invariant ionization frequency per EP.
The ionization energy threshold is scaled in proportion to the energy budget of the $g$ microscopic constituents,
\begin{equation}
    E'_\mathrm{ion} = g\,E_\mathrm{ion},
    \label{eq:ep_eion}
\end{equation}
thereby preserving the ratio of relative kinetic energy to ionization cost.

\subsection{Summary}

Table~\ref{tab:ep_scaling} collects the complete EP parameter mapping.
Together, these transformations preserve the local forces, collisional hardness, ionization statistics, and charging dynamics that govern the macroscopic pressure, temperature, and light emission of the sonoluminescing gas, while reducing the simulated particle count by up to four orders of magnitude relative to the physical system.

\begin{table}[h]
\centering
\caption{Ensemble-particle scaling rules. Primed quantities are EP values; unprimed are microscopic (physical) values; $g = N_{\mathrm{real}}/N_{\mathrm{ensem}}$ is the enlargement factor.}
\label{tab:ep_scaling}
\begin{tabular}{@{}lll@{}}
\toprule
Category & Quantity & Scaling rule \\
\midrule
Momentum   & Mass                & $m' = g\,m$ \\
                  & Velocity            & $v' = v$ \\
\addlinespace
LJ potential      & Diameter            & $\sigma' = g^{1/3}\sigma$ \\
                  & Well depth          & $\epsilon' = g\,\epsilon$ \\
                  & Cutoff radius       & $r_c' = g^{1/3}r_c$ \\
\addlinespace
DSF Coulomb       & Cutoff radius       & $r_c' = g^{1/3}r_c$ \\
                  & Damping parameter   & $\alpha' = g^{-1/3}\alpha$ \\
                  & Charge              & $q_i' = q_i$ \\
                  & Force scale factor  & $s_{\mathrm{pair}} = g$ \\
\addlinespace
Ionization        & Reactive radius     & $r_{\mathrm{ion}}' = g^{1/2}r_{\mathrm{ion}}$ \\
                  & Energy threshold    & $E_{\mathrm{ion}}' = g\,E_{\mathrm{ion}}$ \\
\bottomrule
\end{tabular}
\end{table}
\section{Simulation Procedures}
\label{sec:sim_proc}

\subsection{Physical setup}

All simulations in this work use argon; Table~\ref{tab:ionization} also lists helium and xenon for reference. The successive ionization potentials are drawn from the NIST Atomic Spectra Database~\cite{nist_asd} and are applied directly as the ionization thresholds in the collision-energy criterion described in Section~\ref{sec:md}.

\begin{table}[h]
    \centering
    \caption{Successive ionization potentials (MJ/mol). Each entry is the energy required to remove one additional electron from the indicated charge state.}
    \label{tab:ionization}
    \begin{tabular}{@{}lcccccccc@{}}
        \toprule
        Gas & \multicolumn{8}{c}{Charge state} \\
        \cmidrule(l){2-9}
         & Neutral & 1+ & 2+ & 3+ & 4+ & 5+ & 6+ & 7+ \\
        \midrule
        He & 2.37 & 5.25 &  &  &  &  &  &  \\
        Ar & 1.52 & 2.67 & 3.93 & 5.77 & 7.24 & 8.78 & 12.0 & 13.8 \\
        Xe & 1.17 & 2.05 & 3.10 & 4.60 & 5.76 & 6.93 & 9.46 & 10.8 \\
        \bottomrule
    \end{tabular}
\end{table}

The physical molecule count inside a sonoluminescing argon bubble is obtained from the ideal gas law:
\begin{equation}
P_{\infty} \left(\frac{4 \pi}{3} R_0^3\right) = N_{\mathrm{real}}\, k_B T_{\infty}.
\end{equation}
For the setup used throughout this paper ($R_0 = 4.5\,\mu\mathrm{m}$, $T_{\infty} = 300\,\mathrm{K}$, $P_{\infty} = 1\,\mathrm{atm}$), this gives $N_{\mathrm{real}} \approx 9.3 \times 10^9$ molecules, which is computationally challenging to simulate at full atomistic resolution.
We therefore employ the ensemble-particle (EP) representation described in Section~\ref{sec:ep_scaling}, where each EP stands in for $g = N_{\mathrm{real}}/N_{\mathrm{ensem}}$ microscopic atoms.
Production simulations use ensemble sizes $N_{\mathrm{ensem}} \in \{10^6, 10^7, 10^8\}$, corresponding to EP diameters of $21.1\,d$, $9.8\,d$, and $4.5\,d$ respectively (with $d = 3.66\,\text{\AA}$ for argon), and behavior with respect to $N_{\mathrm{ensem}}$ is examined in Section~\ref{sec:results_scale}.

\begin{table}[h]
    \centering
    \caption{Argon force-field and ionization-cutoff parameters. The LJ and geometric values are physical (single-atom) quantities; ionization cutoffs are given at the EP scale and depend on $g = N_\mathrm{real}/N_\mathrm{ensem}$.}
    \label{tab:params}
    \begin{tabular}{@{}llll@{}}
        \toprule
        Category & Parameter & Symbol & Value \\
        \midrule
        LJ potential & Length scale  & $\sigma$            & $3.401\,\text{\AA}$ \\
                     & Well depth    & $\varepsilon/k_B$   & $116.81\,\text{K}$  \\
        \addlinespace
        Geometry     & Atomic diameter & $d$               & $3.66\,\text{\AA}$  \\
        \addlinespace
        Ionization   & Short-range cutoff & $r_\mathrm{ion}$ & $2\,g^{1/3}d\ (=r_\mathrm{LJ})$ \\
        cutoff       & Full-range cutoff  & $r_\mathrm{ion}$ & $g^{1/2}d$ \\
        \bottomrule
    \end{tabular}
\end{table}

\subsection{Simulation time window and initialization}
\label{sec:init}

Prior EBMD studies of sonoluminescence initialize the bubble at its maximum radius and simulate the entire inward collapse, which is computationally feasible in event-based codes because computation is proportional to the number of collision events rather than wall-clock time.
TBMD, by contrast, advances every timestep regardless of activity; simulating a full acoustic half-cycle at the sub-femtosecond resolution required near collapse would be prohibitive.
We therefore initialize the simulation with the bubble already contracting and restrict the simulation to the 10~ns fine-resolution window identified in Section~\ref{sec:numerical_solver}.
The continuum solution provides the bubble radius, wall velocity, and boundary temperature at the window start time, which are used directly to initialize the MD system.
This starting point is justified by the prior finding that both uniform and nonuniform continuum theories remain accurate through the pre-collapse approach, so that the MD need only resolve the final compression and rebound where the two descriptions diverge.

Particle positions are initialized to reproduce the radial density profile $\rho(r)$ from the continuum solution.
Particles are placed on a cubic lattice with spacing $2d_{\mathrm{ensem}}$ spanning the bubble volume, where $d_{\mathrm{ensem}} = g^{1/3} d$ is the ensemble-particle diameter, $d = 3.66\,\text{\AA}$ is the argon atomic diameter, and $g = N_{\mathrm{real}}/N_{\mathrm{ensem}}$; sites are accepted or rejected in parallel according to the local value of $\rho(r)$, yielding a collision-free initial configuration for all ensemble sizes without pairwise overlap checks.
Initial velocities are drawn from the Maxwell--Boltzmann distribution at the local temperature $T(r)$ predicted by the continuum model, with isotropically randomized directions; the bulk inward velocity field $u_g = (\dot{R}_b/R_b)r$ is not superimposed at $t_0$.
The inward-moving wall naturally imparts this bulk momentum to the gas through particle--wall collisions during the pre-collapse window, so no explicit correction is needed at initialization.

\subsection{Diagnostics and post-processing}
\label{sec:diagnostics}

Macroscopic flow quantities are obtained by averaging particle data within spherical shells of equal volume.
The radial profiles of bulk velocity, temperature, density, and pressure are computed as
\begin{align}
u_g(r) &= \frac{1}{N_r} \sum_{i=1}^{N_r} \frac{\bm{v}_i \cdot \bm{r}_i}{|\bm{r}_i|}, \notag\\
T_b(r) &= \frac{m}{3 k_B N_r} \sum_{i=1}^{N_r} \left|\bm{v}_i - u_g(r)\hat{\bm{r}}_i\right|^2, \notag\\
\rho_b(r) &= \frac{N_r}{V_r}, \notag\\
P_b(r) &= \rho_b(r) k_B T_b(r) + \frac{1}{3 V_r} \sum_{i=1}^{N_r} \bm{r}_i \cdot \bm{f}_i,
\label{eq:MD_properties}
\end{align}
where $N_r$ is the number of particles in the shell between $r-\delta r$ and $r+\delta r$, $V_r$ is the shell volume, $m$ is the microscopic atomic mass (not the EP mass $m'=gm$), and $\bm{f}_i$ is the total force on particle $i$ (the last term is the virial pressure contribution).
The instantaneous bubble radius and wall velocity are advanced by the Keller--Miksis equation (Eq.~\eqref{eq:PDE:ddRb}), driven at each timestep by the gas pressure $P_b$ computed from the MD particle data; the liquid boundary-layer thickness $\delta(t)$ is prescribed by the continuum solution (Eq.~\eqref{eq:PDE:delta}), and all other quantities evolve self-consistently from the particle dynamics.

Raw per-bin diagnostics are subject to Poisson shot noise whenever a shell contains few particles, most severely near the bubble center at early times (small shell volume) and near the wall after rebound (gas rarefaction).
To suppress this noise, we merge consecutive bins adaptively from the outermost shell inward until each merged group contains at least $N_{\mathrm{min}} = 100$ ensemble particles, and assign each group the group-averaged temperature, density, and pressure.
All particle-count and density fields are then rescaled by the factor $g = N_{\mathrm{real}}/N_{\mathrm{ensem}}$ to convert from simulated to physical particle numbers; this rescaling applies to the number-density term in $P_b$ only, as the virial term is already physically correct through the $g$-scaled EP forces.
The scalar peak temperature $T_{\mathrm{max}}$ reported in the results tables is evaluated only over bins whose rescaled count satisfies $N_{\mathrm{count}} \geq 10$, excluding wall-adjacent bins that contain too few physical particles to yield a statistically meaningful estimate.
Space--time temperature heatmaps use a robust display cap at the $99.9$th percentile of the non-empty-bin temperature distribution, which suppresses extreme outlier cells while preserving the interior radial structure.

\section{Results}
\label{sec:results}

We structure the results as follows.
Section~\ref{sec:results_scale} considers behavior with respect to ensemble particle number $N_\mathrm{ensem}$ and validates the ensemble-particle (EP) scaling derived in Section~\ref{sec:ep_scaling}.
Section~\ref{sec:results_boundary} investigates the sensitivity to the thermal accommodation coefficient $\alpha_t$, which governs the gas--liquid energy exchange at the bubble wall.
Section~\ref{sec:results_pair} examines the role of pair-interaction physics, comparing simulations with no ionization, short-range ionization only, and full-range ionization with Coulomb forces.
Section~\ref{sec:results_compare} places our results in the context of prior simulation work through a comprehensive parameter-sweep table.

% -----------------------------------------------------------------------
\subsection{Trends with increasing ensemble particle number}
\label{sec:results_scale}
% -----------------------------------------------------------------------

A key question motivating our study is: how many particles is sufficient to effectively have a ``fully-resolved'' SBSL simulation?
In other words, do we observe convergent or similar behavior as the number of EPs is increased?
From another perspective, we wish to validate our EP model and scaling factors by determining whether macroscopic observables behave similarly as $N_\mathrm{ensem}$ increases.
We vary $N_\mathrm{ensem} \in \{10^6,\,10^7,\,10^8\}$ while holding all other parameters fixed ($\alpha_t = 1.0$, full-range ionization with Coulomb forces), spanning two orders of magnitude and reaching $10^8$ particles---to our knowledge the largest sonoluminescence MD simulation performed to date.

\begin{figure}[htbp]
  \centering
  \subfloat[]{\includegraphics[width=0.33\textwidth]{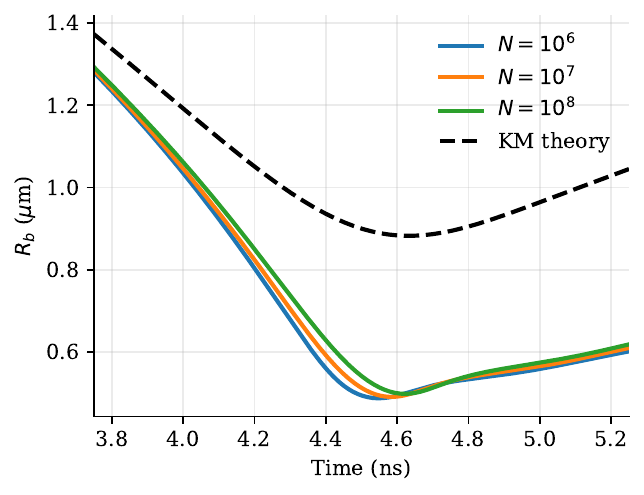}\label{fig:scaling-global-radius}}
  \subfloat[]{\includegraphics[width=0.33\textwidth]{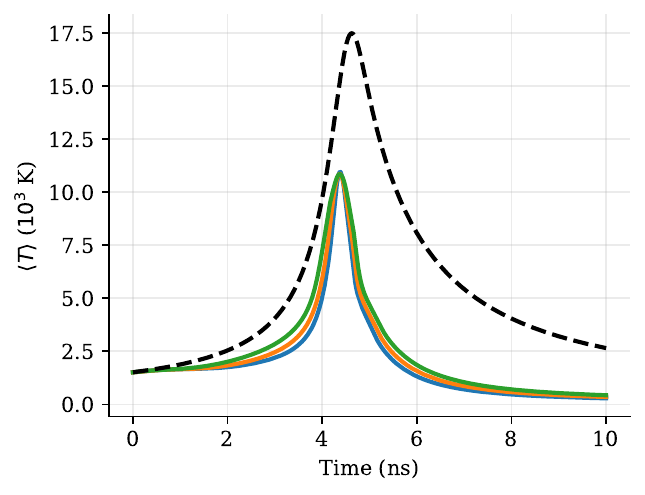}\label{fig:scaling-global-avgT}}
  \subfloat[]{\includegraphics[width=0.33\textwidth]{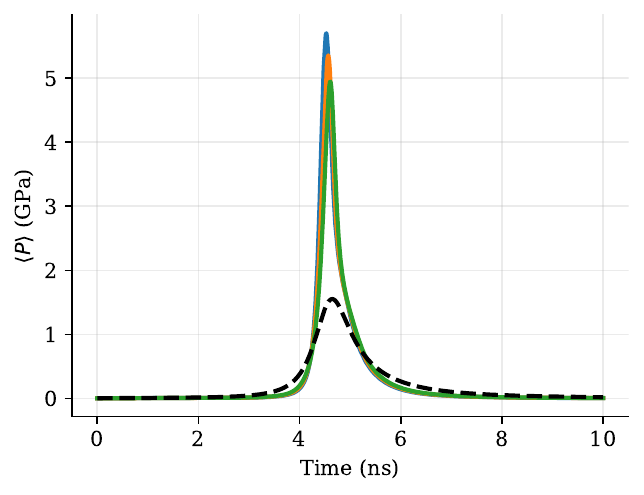}\label{fig:scaling-global-avgP}}
  \caption{Global observables across ensemble particle numbers ($\alpha_t=1.0$, full-range ionization with Coulomb forces).
    (a) Bubble radius $R(t)$.
    (b) Volume-averaged gas temperature $\langle T\rangle(t)$.
    (c) Volume-averaged gas pressure $\langle P\rangle(t)$.
    Colors as in panel~(a) legend; black dashed: Keller--Miksis theoretical prediction.}
  \label{fig:scaling-global}
\end{figure}

Figure~\ref{fig:scaling-global} shows the time evolution of the bubble radius, volume-averaged temperature, and volume-averaged pressure near the collapse point for the three particle counts.
The three MD radius trajectories are mutually consistent, confirming that the ensemble-particle scaling captures the macroscopic collapse kinematics at all three resolutions.
All three share the qualitative collapse shape of the Keller--Miksis prediction, but diverge substantially near the collapse minimum: the MD minimum radius ($R_\mathrm{min} \approx 0.488$--$0.498\,\mu\text{m}$) is roughly half the continuum value ($0.883\,\mu\text{m}$).
Near the collapse minimum, several core assumptions of the Keller--Miksis continuum framework are known to break down simultaneously, including the local thermodynamic equilibrium required for continuum transport, an ideal-gas equation of state with constant $\gamma$, and a thermal conductivity calibrated for $T < 3{,}000\,\mathrm{K}$~\cite{ruuth2002molecular,kim2007molecular,kwak1995aspect}.
The continuum model therefore captures only the gross features of the collapse dynamics and is retained as a qualitative reference; Section~\ref{sec:results_compare} provides quantitative validation of our results against prior MD work.
The time of minimum radius $t_\mathrm{min}$ shifts slightly later as $N_\mathrm{ensem}$ increases (from $4.548\,\text{ns}$ at $10^6$ to $4.622\,\text{ns}$ at $10^8$), and correspondingly the minimum radius $R_\mathrm{min}$ increases modestly from $0.488\,\mu\text{m}$ to $0.498\,\mu\text{m}$.
Both effects are consistent with finer resolution reducing spurious numerical stiffness during the most compressed phase.
The volume-averaged temperature is nearly consistent across the three runs: the peak values are $\langle T\rangle_\mathrm{max} = 10{,}967\,\text{K}$, $10{,}870\,\text{K}$, and $10{,}852\,\text{K}$ for $10^6$, $10^7$, and $10^8$ particles respectively, a variation of less than $1.1\%$ over two orders of magnitude of $N_\mathrm{ensem}$.
The Kwak continuum reference, which neglects ionization, peaks at $17{,}494\,\text{K}$ (approximately $61\%$ above the ionizing MD result), directly attributing the temperature suppression to ionization rather than to ensemble-particle resolution.
The average pressure shows a more pronounced but still systematic trend, with the peak decreasing monotonically from $5.70\,\text{GPa}$ at $10^6$ to $4.95\,\text{GPa}$ at $10^8$ ($\sim$13\% reduction), indicating that lower-resolution runs slightly overestimate the sharpness of the compressive collapse.
The Kwak continuum volume-averaged pressure peaks at $1.55\,\text{GPa}$ (distinct from the wall pressure $p_{W,\mathrm{max}}=1.488\,\text{GPa}$ listed in Table~\ref{tab:compare}), substantially below the MD values; this gap reflects the shallower Kwak collapse ($R_\mathrm{min} = 0.883\,\mu\text{m}$ versus $0.49$--$0.50\,\mu\text{m}$ in MD) rather than a difference in thermodynamic modeling.

\begin{figure}[htbp]
  \centering
  \subfloat[]{\includegraphics[width=0.33\textwidth]{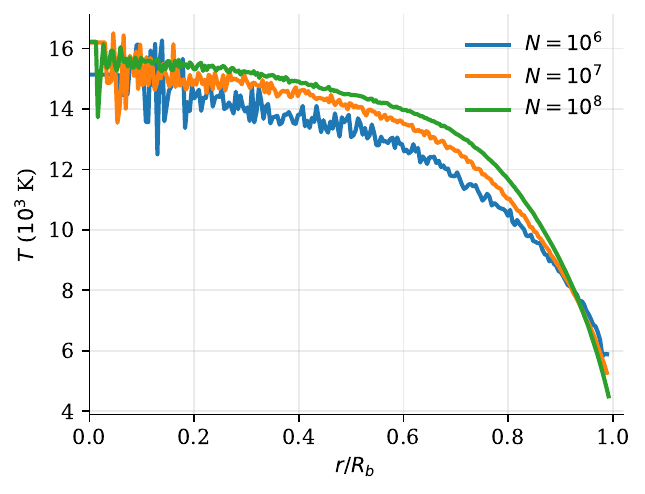}\label{fig:scaling-radial-T}}
  \subfloat[]{\includegraphics[width=0.33\textwidth]{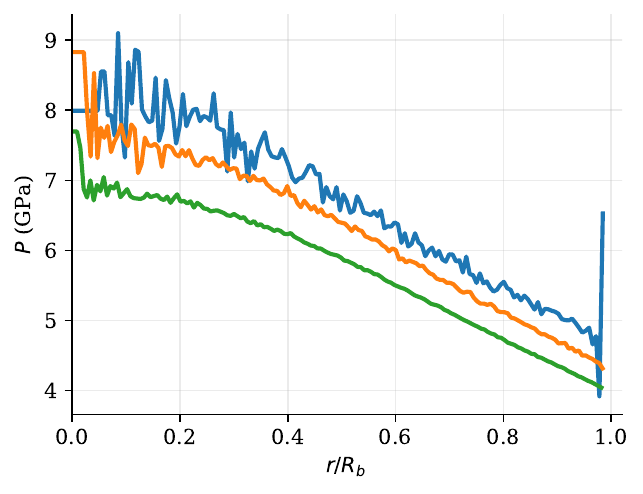}\label{fig:scaling-radial-P}}
  \subfloat[]{\includegraphics[width=0.33\textwidth]{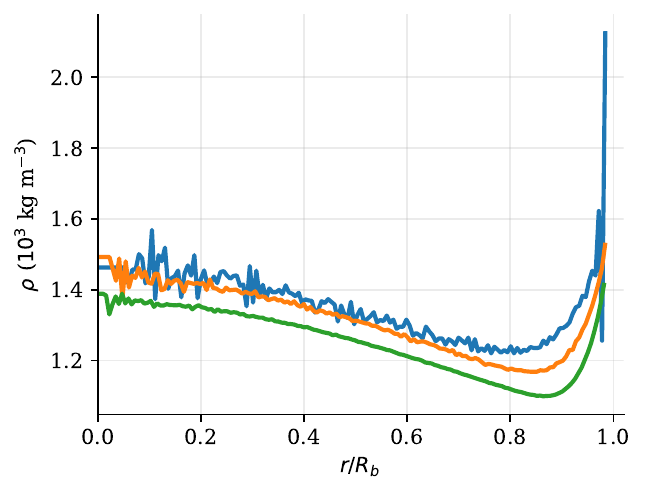}\label{fig:scaling-radial-rho}}
  \caption{Radial profiles inside the bubble at collapse ($\alpha_t=1.0$, full-range ionization with Coulomb forces).
    (a) Temperature $T(r/R_b)$ evaluated at the time of maximum center temperature $t(T_{\mathrm{center,max}})$,
    where the center temperature is defined as the average over the innermost core $r \leq R_b/10$.
    (b) Pressure $P(r/R_b)$ and (c) mass density $\rho(r/R_b)$, both evaluated at minimum bubble radius $t_\mathrm{min}$.
    The outermost two radial bins are excluded because they are contaminated by the gas--wall interaction cutoff.
    Colors as in panel~(a) legend.}
  \label{fig:scaling-radial}
\end{figure}

Figure~\ref{fig:scaling-radial} shows the radial profiles of temperature, pressure, and mass density as a function of the fractional radius $r/R_b$.
The temperature profile (panel~a) is evaluated at the time of maximum center temperature $t(T_{\mathrm{center,max}})$, where the center is defined as the region $r \leq R_b/10$; this snapshot captures the sharpest hot-core structure, which occurs slightly before minimum radius.
The pressure and density profiles (panels b--c) are evaluated at minimum bubble radius $t_\mathrm{min}$, where the compression is greatest.
All three profiles share the same qualitative structure across resolutions: a hot, dense, high-pressure interior that decreases toward the bubble wall.
The temperature profiles display a pronounced hot core: the center temperature (averaged over $r \leq R_b/10$ at $t(T_{\mathrm{center,max}})$) reaches $15{,}536\,\text{K}$, $15{,}354\,\text{K}$, and $15{,}555\,\text{K}$ for $N_\mathrm{ensem} = 10^6$, $10^7$, and $10^8$ respectively, occurring at $0.266\,\text{ns}$, $0.346\,\text{ns}$, and $0.431\,\text{ns}$ before minimum radius.
The center temperature does not converge monotonically with $N_\mathrm{ensem}$, reflecting the inherently higher statistical noise in the small inner bins.
The radial pressure and density profiles are more stable across resolutions: the overall shape and interior magnitude are consistent, with the $10^8$ case yielding slightly lower and smoother profiles owing to better statistical sampling of each radial shell.
Taken together, the behavior in Figs.~\ref{fig:scaling-global}--\ref{fig:scaling-radial} supports the conclusion that the ensemble-particle scaling developed in Section~\ref{sec:ep_scaling} successfully preserves the dominant mesoscopic physics at all three resolutions, and that $N_\mathrm{ensem} = 10^8$ provides the best-resolved reference for quantitative comparisons.

% -----------------------------------------------------------------------
\subsection{Effect of thermal boundary condition}
\label{sec:results_boundary}
% -----------------------------------------------------------------------

The thermal accommodation coefficient $\alpha_t$ governs the rate of energy exchange between gas molecules and the bubble wall upon collision: $\alpha_t = 0$ corresponds to a fully adiabatic wall (specular reflection, no energy transfer), while $\alpha_t = 1$ corresponds to a fully isothermal wall (complete thermalization to the local wall temperature).
Physical estimates for $\alpha_t$ in SL systems are not directly accessible from experiment; Schanz et al.~\cite{dabh10} adopt $\alpha_t = 0.3$ as a representative value, while Kim et al.~\cite{kim2007molecular} consider both limiting cases.
We compare $\alpha_t \in \{0.0,\,0.5,\,1.0\}$ at fixed $N_\mathrm{ensem} = 10^6$ with full-range ionization, to quantify how strongly the choice of $\alpha_t$ shifts the predicted observables.

\begin{figure}[htbp]
  \centering
  \subfloat[]{\includegraphics[width=0.33\textwidth]{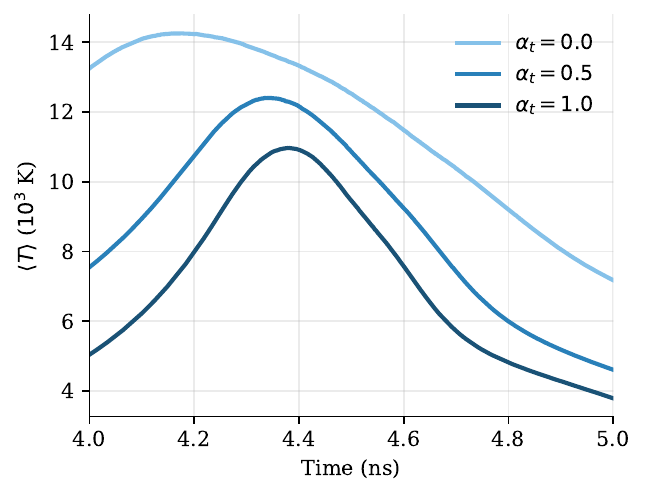}\label{fig:boundary-avgT}}
  \subfloat[]{\includegraphics[width=0.33\textwidth]{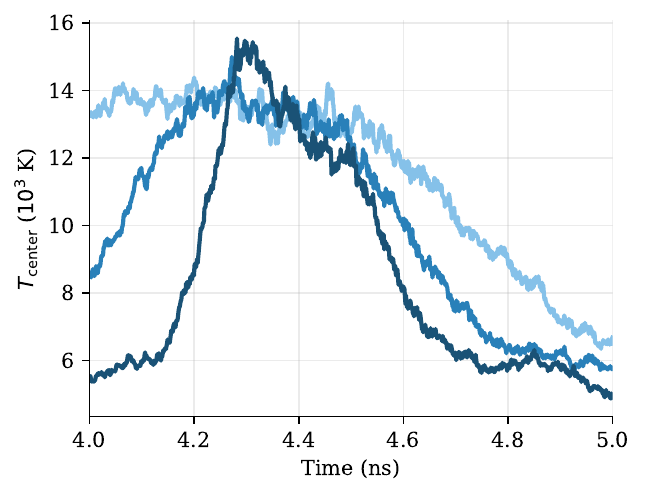}\label{fig:boundary-Tcenter}}
  \subfloat[]{\includegraphics[width=0.33\textwidth]{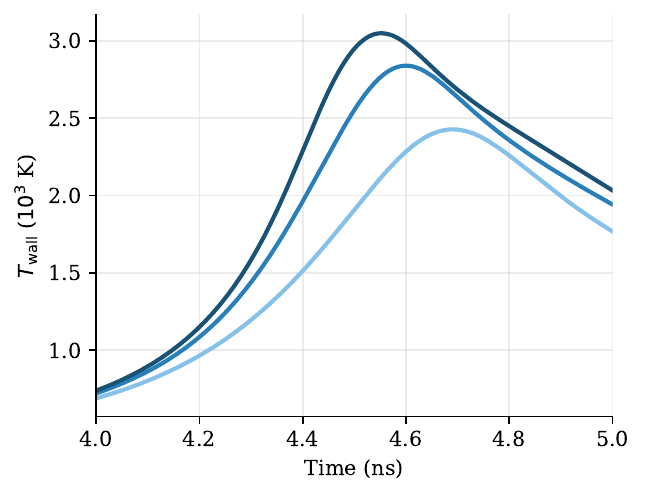}\label{fig:boundary-Twall}}
  \caption{Effect of thermal accommodation coefficient on temperature evolution
    ($N_\mathrm{ensem}=10^6$, full-range ionization with Coulomb forces).
    (a) Volume-averaged temperature $\langle T\rangle(t)$.
    (b) Center temperature $T_\mathrm{center}(t)$.
    (c) Wall temperature $T_\mathrm{wall}(t)$.
    Colors as in panel~(a) legend.}
  \label{fig:boundary-timeseries}
\end{figure}

Figure~\ref{fig:boundary-timeseries} shows the time evolution of the volume-averaged temperature, center temperature, and wall temperature for the three accommodation coefficients.
The volume-averaged temperature, shown in panel~(a), decreases monotonically with increasing $\alpha_t$: the peak values are $14{,}257\,\text{K}$, $12{,}404\,\text{K}$, and $10{,}967\,\text{K}$ for $\alpha_t = 0.0$, $0.5$, and $1.0$ respectively.
This monotone reduction reflects the increasing rate at which energy is transferred from the gas to the liquid as the accommodation coefficient grows.
More strikingly, panels~(b) and~(c) reveal a physically distinct behavior at the bubble center and wall.
The center temperature exhibits the opposite ordering: $T_{\mathrm{center,max}} = 14{,}380\,\text{K}$, $14{,}989\,\text{K}$, and $15{,}536\,\text{K}$ for $\alpha_t = 0.0$, $0.5$, and $1.0$.
That is, stronger wall coupling produces a \textit{hotter} center despite a lower global average.
The wall temperature follows the same trend as the center, increasing from $2{,}429\,\text{K}$ to $3{,}050\,\text{K}$ as $\alpha_t$ increases.
This behavior is physically consistent with the gas--wall energy exchange model of Section~\ref{sec:md}: a larger $\alpha_t$ increases the heat flux $P_Q$ transferred from the hot gas to the liquid shell at each wall collision, which raises the liquid-side boundary temperature $T_{bl}$ through the shell energy update.
Since gas molecules adjacent to the wall are thermalized toward this higher $T_{bl}$ upon collision, the measured gas-side wall temperature rises in tandem.
A further consequence of increasing $\alpha_t$ is a systematic advance of the collapse: the time of minimum radius shifts from $t_\mathrm{min} = 4.720\,\text{ns}$ at $\alpha_t = 0$ to $4.548\,\text{ns}$ at $\alpha_t = 1.0$, and the thermal peak moves progressively closer in time to the radius minimum.
At $\alpha_t = 0.0$ the average temperature peaks approximately $0.55\,\text{ns}$ before $t_\mathrm{min}$; at $\alpha_t = 1.0$ this interval narrows to only $0.17\,\text{ns}$, indicating that stronger wall thermalization sharpens and delays the thermal concentration toward the final collapse stage.

\begin{figure}[htbp]
  \centering
  \subfloat[$\alpha_t=0.0$]{\includegraphics[width=0.33\textwidth]{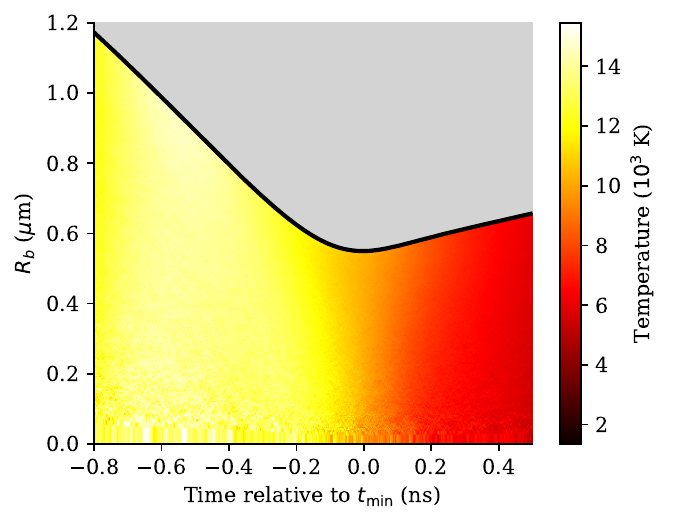}\label{fig:boundary-heatmap-0}}
  \subfloat[$\alpha_t=0.5$]{\includegraphics[width=0.33\textwidth]{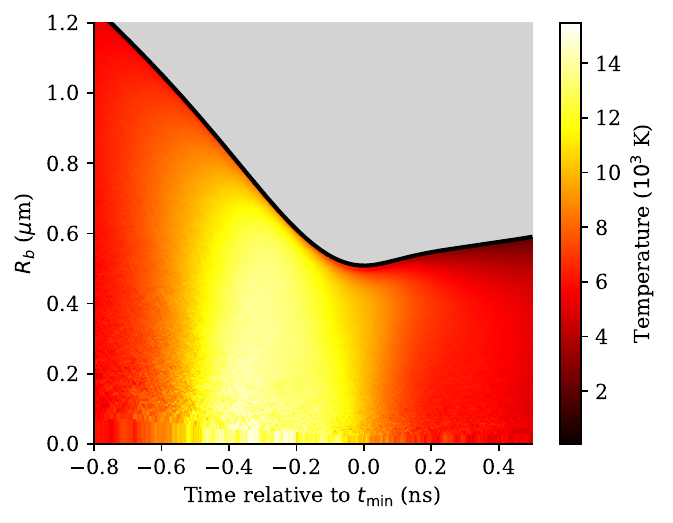}\label{fig:boundary-heatmap-05}}
  \subfloat[$\alpha_t=1.0$]{\includegraphics[width=0.33\textwidth]{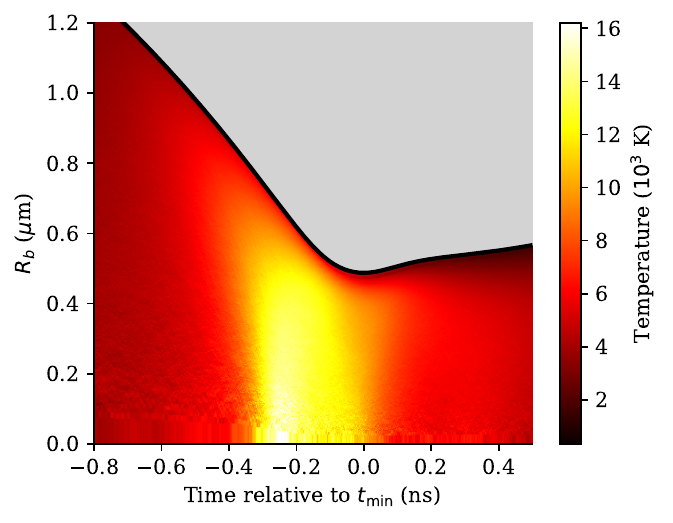}\label{fig:boundary-heatmap-1}}
  \caption{Space--time evolution of gas temperature $T(r,\,t)$ for different thermal accommodation coefficients
    ($N_\mathrm{ensem}=10^6$, full-range ionization with Coulomb forces).
    The horizontal axis is simulation time; the vertical axis is the absolute bubble radius $r\,(\mu\mathrm{m})$.
    The instantaneous bubble boundary $R_b(t)$ is overlaid as a solid black curve; the exterior region is shaded.
    Color encodes temperature in units of $10^3\,\text{K}$.
    The time of minimum radius for each case is $t_\mathrm{min} = 4.720\,\text{ns}$ ($\alpha_t=0.0$), $4.603\,\text{ns}$ ($\alpha_t=0.5$), and $4.548\,\text{ns}$ ($\alpha_t=1.0$); these values can be used to align each heatmap with the corresponding time-series in Fig.~\ref{fig:boundary-timeseries}.
    Each panel uses an independent robust display cap ($99.9$th percentile of the occupied-bin distribution) to preserve the interior radial texture.}
  \label{fig:boundary-heatmaps}
\end{figure}

The apparently paradoxical inversion between the average and center temperature trends is resolved by examining the spatial temperature structure.
Figure~\ref{fig:boundary-heatmaps} shows the space--time evolution of the gas temperature inside the bubble for each boundary condition, with the instantaneous bubble radius overlaid.
At $\alpha_t = 0.0$ (panel~a), the temperature field is relatively broad and spatially uniform: energy is retained throughout the gas, the average is high, but the center does not develop a strong concentration relative to the bulk.
As $\alpha_t$ increases, the wall continuously extracts energy from the outer gas layers, steepening the radial thermal gradient.
At $\alpha_t = 1.0$ (panel~c), the result is a pronounced hot core coexisting with a relatively cool near-wall region: energy is concentrated inward rather than distributed uniformly.
The total thermal energy budget of the gas is reduced (lower average temperature) but the spatial focusing is stronger, driving the center to higher absolute temperatures than in the adiabatic case.
The intermediate case $\alpha_t = 0.5$ (panel~b) bridges these extremes cleanly, displaying an intermediate degree of radial stratification.
Since the primary radiative source in SL is the hot central plasma, these results indicate that the isothermal boundary condition, despite lowering the bulk temperature, produces the most localized and intense light-emitting region.
The choice of $\alpha_t$ therefore has qualitative as well as quantitative consequences for the predicted light emission, and experimental calibration of this parameter is important for reliable emission estimates.

% -----------------------------------------------------------------------
\subsection{Effect of pair interactions and ionization}
\label{sec:results_pair}
% -----------------------------------------------------------------------

Prior MD studies of sonoluminescence have modeled gas particles as either hard spheres~\cite{dabh10,ruuth2002molecular} or with a purely short-range Lennard--Jones potential~\cite{kim2007molecular}, neglecting both the energy cost of ionization events and the long-range Coulomb interaction between the resulting charge carriers.
At the extreme temperatures reached during SL collapse, however, significant ionization is expected: for argon under the conditions studied here, the Saha equilibrium predicts appreciable ionization fractions above approximately $15{,}000\,\text{K}$.
We compare three levels of pair-interaction physics at fixed $N_\mathrm{ensem} = 10^6$ and $\alpha_t = 1.0$:
\emph{no ionization} (LJ potential only), \emph{short-range ionization}, and \emph{full-range ionization}.
Both ionized configurations use the identical Lennard--Jones and DSF Coulomb interactions and differ only in the ionization reactive cutoff $r_\mathrm{ion}$ (see Table~\ref{tab:params}): the short-range case sets $r_\mathrm{ion} = r_\mathrm{LJ} = 2\,g^{1/3}d$, confining ionization events to within the LJ interaction range, while the full-range case extends it to $r_\mathrm{ion} = g^{1/2}d$, allowing ionization over a wider separation.
We note that ionization events in the simulation are treated as irreversible; the light-emission estimates below are therefore obtained as a post-processing diagnostic by applying the Saha--LTE equation \cite{zel2002physics} to the local MD temperature and number-density fields, following the approach of Schanz et al.~\cite{dabh10}.
In the strongly-coupled plasma regime reached near collapse, ionization-potential lowering would increase the true ionization fraction beyond these Saha estimates~\cite{khalid2012opacity}; the emission values below should therefore be interpreted as order-of-magnitude lower bounds rather than quantitative predictions.

\begin{figure}[htbp]
  \centering
  \subfloat[]{\includegraphics[width=0.33\textwidth]{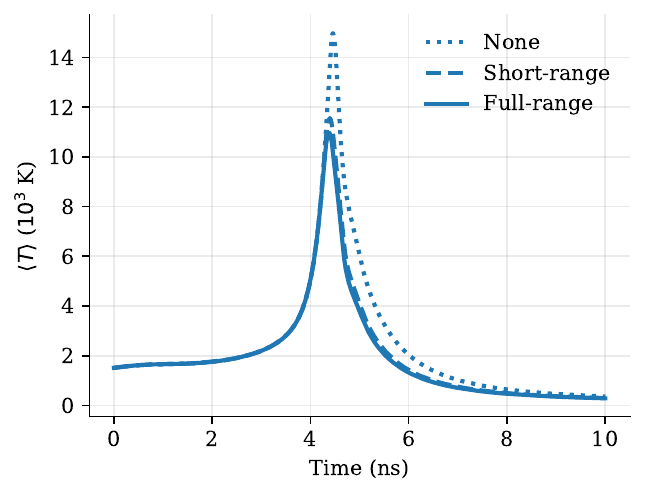}\label{fig:pair-avgT}}
  \subfloat[]{\includegraphics[width=0.33\textwidth]{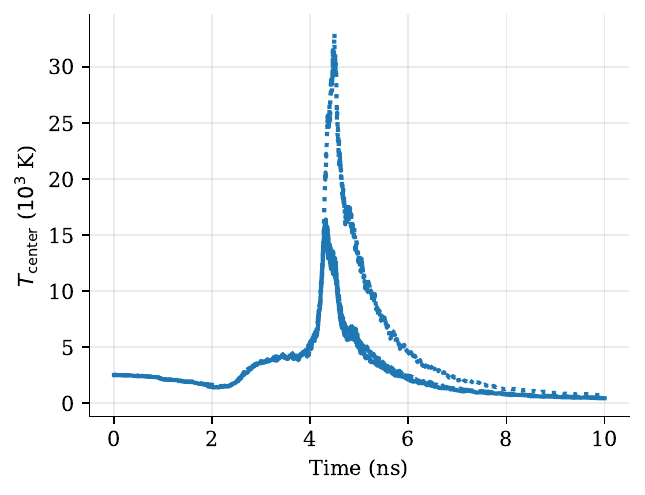}\label{fig:pair-Tcenter}}
  \subfloat[]{\includegraphics[width=0.33\textwidth]{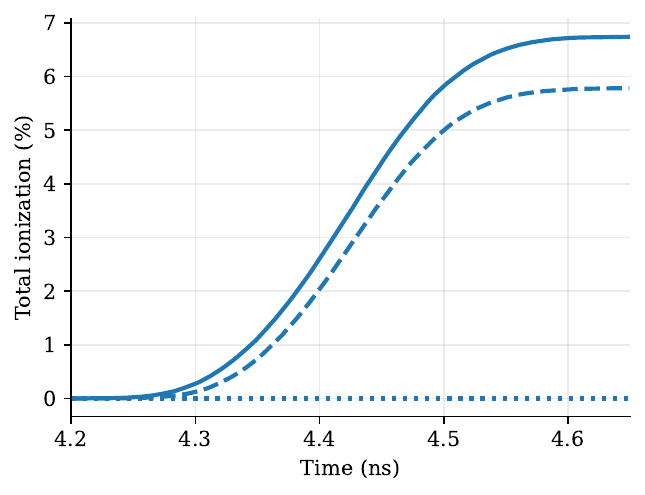}\label{fig:pair-chargesum}}
  \caption{Effect of pair-interaction model on collapse dynamics
    ($N_\mathrm{ensem}=10^6$, $\alpha_t=1.0$).
    (a) Volume-averaged temperature $\langle T\rangle(t)$.
    (b) Center temperature $T_\mathrm{center}(t)$.
    (c) Total simulation charge normalized by $N_\mathrm{ensem}$, serving as a proxy for the instantaneous ionization fraction.
    Colors and line styles as in panel~(a) legend.}
  \label{fig:pair-timeseries}
\end{figure}

Figure~\ref{fig:pair-timeseries} compares the volume-averaged temperature, center temperature, and total simulation charge as functions of time for the three interaction models.
The volume-averaged temperature (panel~a) decreases monotonically with the level of ionization physics included: the peak values are $14{,}975\,\text{K}$ (no ionization), $11{,}557\,\text{K}$ (short-range), and $10{,}967\,\text{K}$ (full-range), representing reductions of $23\%$ and $27\%$ relative to the no-ionization baseline.
The center temperature (panel~b) shows a far more dramatic effect: the no-ionization run reaches $T_\mathrm{center,max} \approx 33{,}000\,\text{K}$, while both ionized runs remain near $15{,}500$--$16{,}400\,\text{K}$ — a suppression of roughly a factor of two.
This large discrepancy arises because, in the absence of any ionization energy sink, kinetic energy at the bubble center accumulates without bound during the final stages of compression, producing unrealistically high local temperatures.
Once ionization is active, the energy consumed by successive electron-removal events acts as a distributed internal energy sink that prevents runaway heating of the hottest gas.
The difference between the two ionized configurations is modest by comparison ($\Delta T_\mathrm{center} \approx 870\,\text{K}$ relative to the short-range case): the larger ionization reactive cutoff $r_\mathrm{ion}$ of the full-range case admits ionization events over a wider separation near the hot core, deepening the ionization energy sink and further lowering the peak center temperature.
Panel~(c) confirms that the net simulation charge remains identically zero in the no-ionization run, while both ionized models accumulate charge during the collapse, reaching peaks of $5.78\%$ (short-range) and $6.74\%$ (full-range) of the ensemble at the time of most intense ionization---which occurs slightly \emph{after} the thermal maximum, as the charge carriers persist into the post-collapse phase.

\begin{figure}[htbp]
  \centering
  \subfloat[No ionization]{\includegraphics[width=0.33\textwidth]{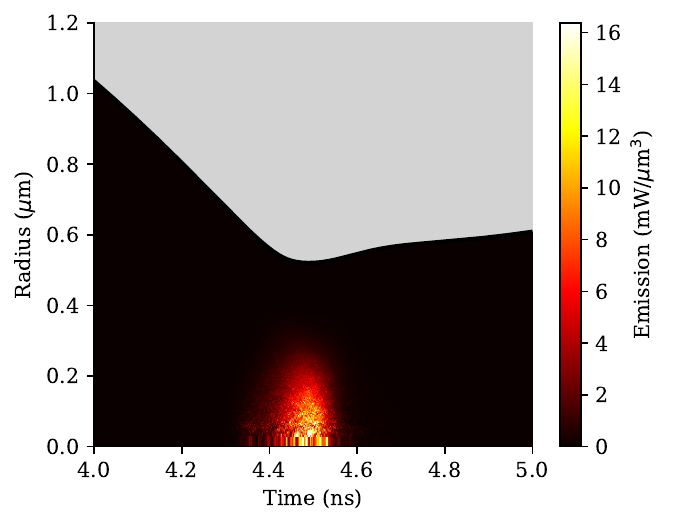}\label{fig:pair-emission-noio}}
  \subfloat[Short-range ionization]{\includegraphics[width=0.33\textwidth]{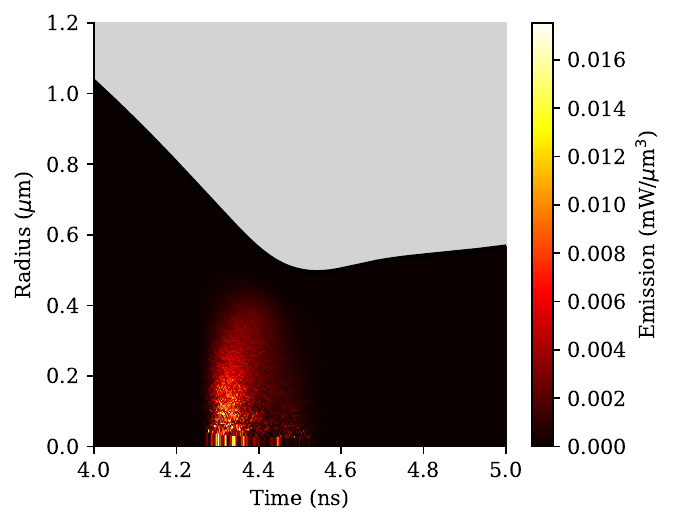}\label{fig:pair-emission-shortio}}
  \subfloat[Full-range ionization]{\includegraphics[width=0.33\textwidth]{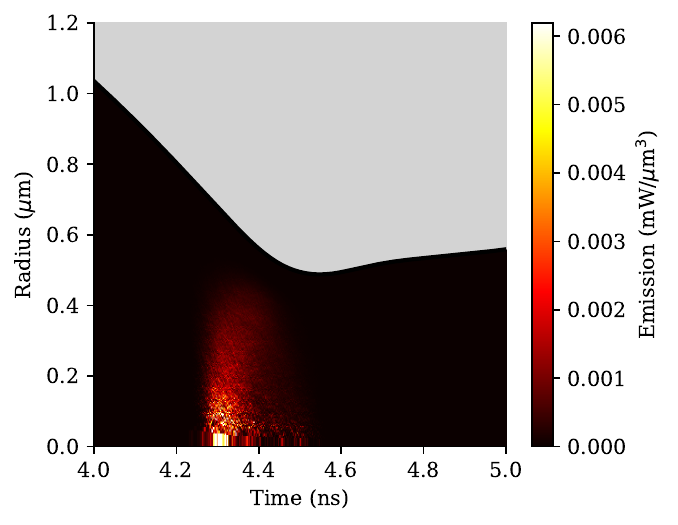}\label{fig:pair-emission-fullio}}
  \caption{Space--time distribution of estimated light emission for three pair-interaction models
    ($N_\mathrm{ensem}=10^6$, $\alpha_t=1.0$).
    Emission is computed as a post-processing diagnostic via a single-stage Saha--LTE ionization model applied to the local MD temperature and number-density fields, following~\cite{dabh10}, and band-limited to wavelengths $\lambda \geq 200\,\text{nm}$.
    Axes and overlay conventions follow Fig.~\ref{fig:boundary-heatmaps}.
    Each panel uses an independent robust display cap; the no-ionization case exceeds the ionized cases by approximately three orders of magnitude in absolute emission, and cross-panel colors are not directly comparable.}
  \label{fig:pair-emission}
\end{figure}

The practical consequence for the predicted light output is illustrated in Fig.~\ref{fig:pair-emission}, which shows the space--time distribution of a Saha--LTE emission estimate (bolometric, band-limited to $\lambda \geq 200\,\text{nm}$) inferred from the local MD temperature and density fields.
The no-ionization run produces a peak emission approximately three orders of magnitude larger than either ionized case.
This dramatic difference arises from two compounding mechanisms.
First, the ionization algorithm described in Section~\ref{sec:md} removes kinetic energy from colliding particles at each ionization event, directly lowering the gas temperature that enters the Saha calculation: the center temperature in the ionized runs ($\sim15{,}500$--$16{,}400\,\text{K}$) is roughly half that of the no-ionization run ($\sim33{,}000\,\text{K}$).
Second, the Saha--LTE ionization fraction scales exponentially with inverse temperature, $q \propto \exp(-\chi/k_BT)$, and the emission proxy scales as $q^2 n^2 T^{1/2}$; a factor-of-two reduction in $T$ therefore produces an orders-of-magnitude reduction in $q$ that enters the emission as $q^2$.
The three-order-of-magnitude drop in emission is thus an amplification of the factor-of-two temperature suppression through the Saha exponential, not a direct proportionality.
To preserve the spatial structure within each panel, the heatmaps use independent per-panel display scales; absolute cross-panel comparison should be made using the scalar totals in Table~\ref{tab:sweep}.
In all three cases the emission is sharply concentrated near the bubble center in a brief temporal window around the collapse.
These results demonstrate that neglecting ionization not only introduces a quantitative bias in the temperature prediction, but leads to emission estimates that are qualitatively inconsistent with physically expected magnitudes.
Extending the ionization reactive cutoff provides a further, physically motivated refinement beyond the short-range ionization treatment alone.

% -----------------------------------------------------------------------
\subsection{Comparison with prior work}
\label{sec:results_compare}
% -----------------------------------------------------------------------

Table~\ref{tab:compare} presents a three-way cross-model comparison of peak observables among this work, the Kwak continuum model, and the Schanz et al.\ EBMD reference, all under identical driving conditions.
Table~\ref{tab:sweep} collects the full eleven-configuration TBMD parameter sweep.
We first discuss scalar peak observables in this subsection; the qualitative structure of the collapse---the relative timing of the velocity and temperature maxima, the radial profiles, and the shape of the emission pulse---is taken up separately in Section~\ref{sec:results_general}.

The reported $T_\mathrm{max}$ in both tables is evaluated over radial bins with at least ten particle counts (after the spatial smoothing described in Section~\ref{sec:diagnostics}), to avoid domination by statistically underpopulated wall-adjacent bins.
The $f_\mathrm{ion}$ column reveals two trends: it decreases with increasing $\alpha_t$ because stronger wall thermalization lowers the collapse temperature available for ionization, and it increases with $N_\mathrm{ensem}$ at fixed physics parameters because finer EP resolution captures the hot-core ionization events more completely.

The Kwak continuum (PDE) row in Table~\ref{tab:compare} uses our implementation of the Kwak \& Na model~\cite{kwak1997physical} solved under the same acoustic and thermophysical conditions as the MD runs.
This model predicts $T_\mathrm{max} = 34{,}927\,\text{K}$ and $R_\mathrm{min} = 0.883\,\mu\text{m}$.
The elevated temperature, arising because the continuum formulation contains no explicit ionization energy sink, is quantitatively on the same order as our no-ionization TBMD runs ($T_\mathrm{max} = 39{,}617$--$45{,}820\,\text{K}$), confirming that the Kwak PDE and our MD approaches predict qualitatively similar collapse extrema when ionization is absent.

Several trends are immediately apparent within the TBMD parameter sweep.
First, our no-ionization runs produce $T_\mathrm{max}$ values
%between $39{,}617\,\text{K}$ and $45{,}820\,\text{K}$,
substantially above the Schanz reference value of $18{,}870\,\text{K}$, confirming that ionization is an essential physical process for obtaining realistic temperature predictions.
The absence of such extreme temperatures in the Schanz et al.\ results, despite their model not incorporating an explicit ionization energy sink during the MD itself, is attributable to two features of their modeling approach.
Their event-based hard-sphere formulation yields a different compressibility and thermal equation of state than our soft-sphere LJ model.
More significantly, their model includes water vapor chemistry: a substantial quantity of $\mathrm{H_2O}$ molecules is trapped in the bubble during collapse and subsequently dissociates, providing an internal energy buffer that prevents extreme center-temperature excursions~\cite{dabh10}.
In our argon-only TBMD model, the explicit collision-based ionization algorithm~\cite{ruuth2002molecular} plays the analogous energy-limiting role, and our ionized configurations consequently bracket the Schanz temperature range closely.
Second, for configurations that include ionization, $T_\mathrm{max}$ falls in the range $17{,}272$--$21{,}265\,\text{K}$, bracketing the Schanz value comfortably.
Our best-resolved result, $N_\mathrm{ensem} = 10^8$ with full-range ionization and $\alpha_t = 1.0$, yields $T_\mathrm{max} = 17{,}796\,\text{K}$ and $T_\mathrm{av,max} = 10{,}852\,\text{K}$.
The $\alpha_t = 0.5$ full-ionization run ($T_\mathrm{av,max} = 12{,}404\,\text{K}$) offers a closer match to the Schanz average temperature, consistent with their adopted value of $\alpha_t = 0.3$ lying between our $0.0$ and $0.5$ cases.
Third, our minimum bubble radii ($R_\mathrm{min} = 0.488$--$0.733\,\mu\text{m}$) are systematically smaller than the Schanz value of $0.750\,\mu\text{m}$.
We attribute this in part to the absence of water vapor in our argon-only model: in Schanz et al., the trapped vapor exerts additional internal pressure that resists the final compression and arrests the collapse at a larger radius.
The maximum wall velocity in our simulations ($936$--$1{,}271\,\text{m\,s}^{-1}$) is correspondingly lower than the Schanz value ($1{,}510\,\text{m\,s}^{-1}$), consistent with this softer rebound.
Wall pressure values are broadly comparable across models.
Overall, the agreement between our ionization-inclusive TBMD results and the EBMD reference is physically reasonable given the different interaction models and the absence of vapor chemistry in the present work.

\begin{table}[htbp]
\centering
\caption{Cross-model comparison of peak observables under identical driving conditions
($R_0=4.5\,\mu\text{m}$, $f=26.5\,\text{kHz}$, $P_A=1.3\,\text{bar}$, pure argon).
$T_\mathrm{max}$: maximum local gas temperature (evaluated over bins with particle count $\geq 10$);
$T_\mathrm{av,max}$: maximum volume-averaged temperature;
$R_\mathrm{min}$: minimum bubble radius;
$v_{W,\mathrm{max}}$: maximum inward wall velocity;
$p_{W,c}$: wall pressure $0.2\,\text{ns}$ before minimum radius;
$p_{W,\mathrm{max}}$: maximum wall pressure.}
\label{tab:compare}
\begin{tabular}{@{}llrrrrrr@{}}
\toprule
Method & $\alpha_t$
  & $T_\mathrm{max}$ & $T_\mathrm{av,max}$ & $R_\mathrm{min}$
  & $v_{W,\mathrm{max}}$ & $p_{W,c}$ & $p_{W,\mathrm{max}}$ \\
 & & (K) & (K) & ($\mu$m) & (m\,s$^{-1}$) & (GPa) & (GPa) \\
\midrule
This work, TBMD ($N_\mathrm{ensem}=10^8$, full) & 1.0 & 17796 & 10852 & 0.498 & 1133 & 1.796 & 3.076 \\
This work, Kwak continuum (PDE)                 & --  & 34927 & 17494 & 0.883 &  728 & 1.247 & 1.488 \\
Schanz et al.~\cite{dabh10}, EBMD               & 0.3 & 18870 & 12269 & 0.750 & 1510 & 0.950 & 3.300 \\
\bottomrule
\multicolumn{8}{p{0.85\textwidth}}{\footnotesize
  Kwak continuum: our implementation of Kwak \& Na~\cite{kwak1997physical} under the same conditions.
  Kim et al.~\cite{kim2007molecular} adopt the same continuum model as their theoretical reference
  but study a different gas--liquid system (noble gas in acid), precluding direct quantitative MD comparison;
  a qualitative discussion is given in Section~\ref{sec:results_general}.} \\
\end{tabular}
\end{table}

\begin{table}[htbp]
\centering
\caption{Peak observables across all eleven TBMD simulation configurations
($R_0=4.5\,\mu\text{m}$, $f=26.5\,\text{kHz}$, $P_A=1.3\,\text{bar}$, pure argon).
Column definitions as in Table~\ref{tab:compare}; $f_\mathrm{ion}$ is the cumulative fraction of ensemble particles ionized over the full simulation (zero by definition for no-ionization runs).}
\label{tab:sweep}
\begin{tabular}{@{}lllrrrrrrr@{}}
\toprule
$N_\mathrm{ensem}$ & $\alpha_t$ & Pair
  & $T_\mathrm{max}$ & $T_\mathrm{av,max}$ & $f_\mathrm{ion}$
  & $R_\mathrm{min}$ & $v_{W,\mathrm{max}}$ & $p_{W,c}$ & $p_{W,\mathrm{max}}$ \\
 & & & (K) & (K) & (\%) & ($\mu$m) & (m\,s$^{-1}$) & (GPa) & (GPa) \\
\midrule
$10^6$ & 0.0 & none  & 45820 & 31825 &  0.0 & 0.733 &  936 & 1.278 & 2.243 \\
$10^6$ & 0.0 & short & 21265 & 15690 & 21.5 & 0.570 &  954 & 1.757 & 2.310 \\
$10^6$ & 0.0 & full  & 18149 & 14257 & 22.6 & 0.550 &  969 & 1.776 & 2.427 \\
$10^6$ & 0.5 & none  & 42577 & 18887 &  0.0 & 0.569 & 1172 & 1.231 & 3.184 \\
$10^6$ & 0.5 & short & 19368 & 13336 &  9.2 & 0.518 & 1171 & 1.927 & 3.051 \\
$10^6$ & 0.5 & full  & 17272 & 12404 & 10.4 & 0.508 & 1171 & 1.918 & 3.080 \\
$10^6$ & 1.0 & none  & 39617 & 14975 &  0.0 & 0.520 & 1270 & 0.966 & 3.797 \\
$10^6$ & 1.0 & short & 19380 & 11557 &  5.8 & 0.494 & 1270 & 1.406 & 3.591 \\
$10^6$ & 1.0 & full  & 18801 & 10967 &  6.7 & 0.488 & 1271 & 1.505 & 3.580 \\
$10^7$ & 1.0 & full  & 18486 & 10870 &  9.2 & 0.491 & 1201 & 1.742 & 3.314 \\
$10^8$ & 1.0 & full  & 17796 & 10852 & 11.5 & 0.498 & 1133 & 1.796 & 3.076 \\
\bottomrule
\end{tabular}
\end{table}

% -----------------------------------------------------------------------
\subsection{Qualitative comparisons}
\label{sec:results_general}

\begin{figure}[htbp]
  \centering
  \subfloat[]{\includegraphics[width=0.33\textwidth]{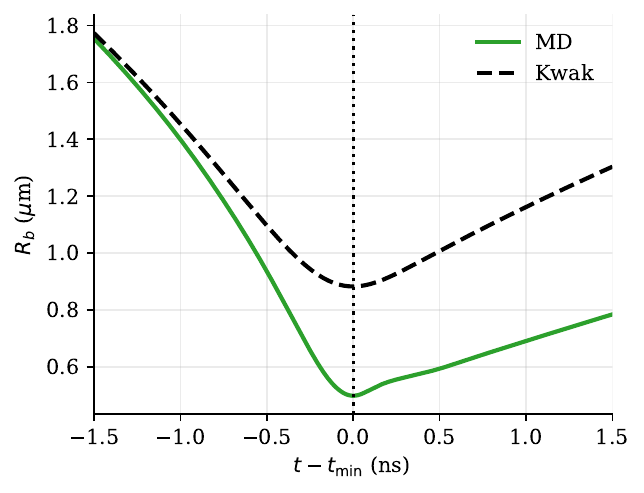}\label{fig:general-dynamics-radius}}
  \subfloat[]{\includegraphics[width=0.33\textwidth]{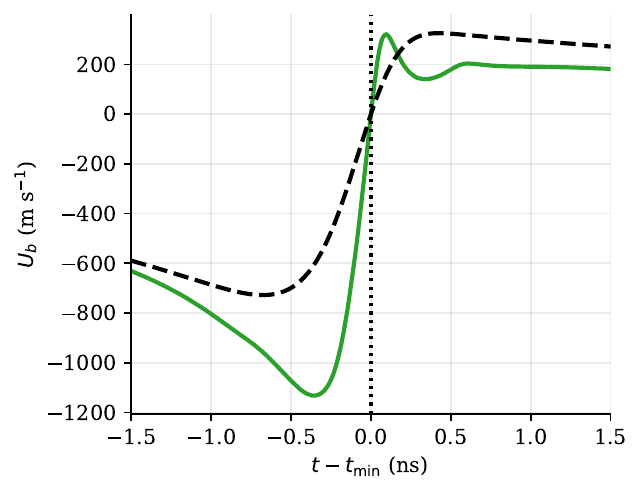}\label{fig:general-dynamics-velocity}}
  \subfloat[]{\includegraphics[width=0.33\textwidth]{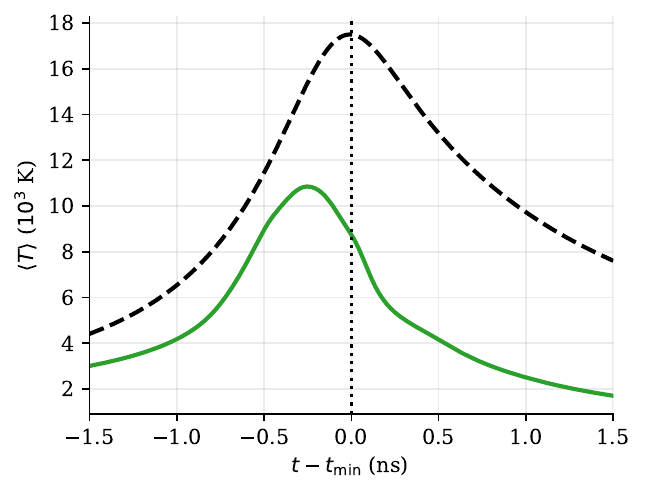}\label{fig:general-dynamics-avgT}}
  \caption{Collapse dynamics for the $N_\mathrm{ensem}=10^8$, full-range ionization, $\alpha_t=1.0$ TBMD run (solid) versus the Kwak continuum PDE solution (dashed), both plotted relative to their own $t_\mathrm{min}$.
    (a) Bubble radius $R(t)$.
    (b) Wall velocity $U_b(t)$; negative values indicate inward motion.
    (c) Volume-averaged gas temperature $\langle T\rangle(t)$.
    Both the wall velocity and the temperature peak before $t_\mathrm{min}$ and fall off at rebound.
    A vertical dotted line marks $t_\mathrm{min}$.}
  \label{fig:general-dynamics}
\end{figure}

\begin{figure}[t]
  \centering
  \subfloat[]{\includegraphics[width=0.33\textwidth]{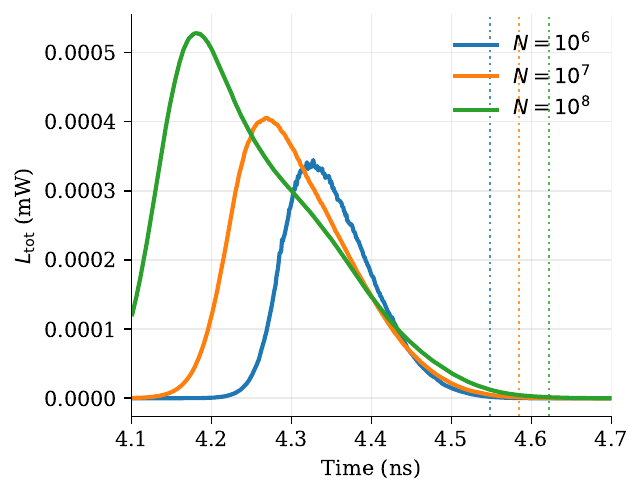}\label{fig:general-emission-total}}
  \subfloat[]{\includegraphics[width=0.33\textwidth]{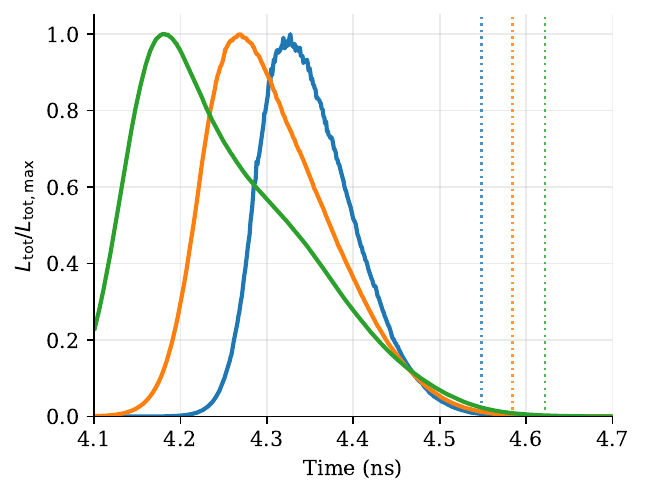}\label{fig:general-emission-norm}}
  \subfloat[]{\includegraphics[width=0.33\textwidth]{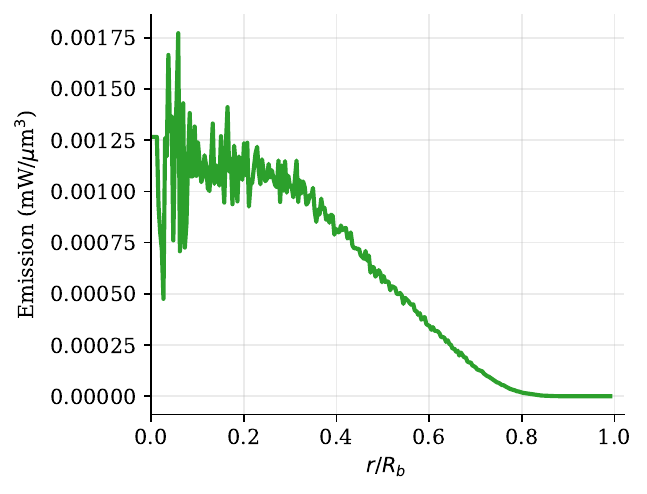}\label{fig:general-emission-radial}}
  \caption{Saha--LTE emission pulse for full-range ionization, $\alpha_t=1.0$, across three ensemble sizes.
    (a) Total emission $L_\mathrm{tot}(t)$ versus time relative to $t_\mathrm{min}$; colors as in panel~(a) legend.
    (b) Same curves normalised to unit peak, showing consistent pulse shape across resolutions.
    (c) Radial emission profile at the time of peak emission ($N_\mathrm{ensem}=10^8$), showing concentration near the bubble centre.
    In all three cases the emission peaks before $t_\mathrm{min}$ (vertical dotted lines in panels~(a) and~(b)) and the pulse width is narrow relative to the acoustic period.}
  \label{fig:general-emission}
\end{figure}

Section~\ref{sec:results_compare} compared scalar peak values; an equally important test of a model is whether it reproduces the \emph{shape} of the collapse. Despite differences in interaction model, gas species, and liquid host, all MD simulations of single-bubble sonoluminescence share a set of universal structural features: the wall velocity and average temperature both reach their maxima before the bubble arrives at its minimum radius, a pronounced hot core develops in the radial thermal structure, and the emission pulse is narrow and concentrated near the centre.
We examine these features in turn through Figures~\ref{fig:general-dynamics}, \ref{fig:scaling-radial}, and~\ref{fig:general-emission}, using our best-resolved TBMD run ($N_\mathrm{ensem}=10^8$, full-range ionization, $\alpha_t=1.0$) and comparing against the prior MD results of Schanz et al.~\cite{dabh10} and Kim et al.~\cite{kim2007molecular}. We do not revisit the scalar comparison of Table~\ref{tab:compare}; instead we reuse the standalone Kwak continuum solution introduced there as a shared baseline.
%(Kwak is a corresponding author),
Because Kim et al.\ adopt the same Kwak \& Na model~\cite{kwak1997physical} as their theoretical reference, this baseline links all three MD frameworks, even though the differing gas--liquid system of Kim et al.\ permits only the qualitative comparison made here.

\paragraph{Collapse dynamics}
Figure~\ref{fig:general-dynamics} shows the bubble radius, wall velocity, and volume-averaged temperature over the final $1.5\,\text{ns}$ before and after minimum radius, for both the TBMD run and the standalone Kwak PDE solution (the Keller--Miksis wall driven by the Kwak continuum gas model rather than by the MD gas pressure), plotted relative to each model's own $t_\mathrm{min}$.
The radius trajectories agree closely, confirming that the KM-coupled continuum model reproduces the large-scale collapse kinematics faithfully.
The wall velocity (panel~b) peaks inward before $t_\mathrm{min}$ and reverses sign at rebound; the corresponding peak inward speeds (TBMD $1{,}133\,\text{m\,s}^{-1}$, Kwak PDE $728\,\text{m\,s}^{-1}$) are the values listed in Table~\ref{tab:compare}.
The volume-averaged temperature (panel~c) also peaks before $t_\mathrm{min}$ in both the MD and PDE solutions.
This pre-minimum timing of both the velocity and temperature maxima is a universal feature of all MD SL simulations: it reflects the fact that the gas heats most strongly when the inward wall speed is greatest, which occurs during the final inward rush rather than at the moment of maximum compression~\cite{dabh10,kim2007molecular}.

\paragraph{Radial structure at collapse}
The radial profiles already shown in Fig.~\ref{fig:scaling-radial} also illustrate this universal collapse structure.
All three runs share the same center-peaked temperature structure (panel~a): the hot core reaches approximately $15{,}000$--$16{,}000\,\text{K}$ and decreases to roughly $4{,}500$--$6{,}000\,\text{K}$ at the outer edge, with the $N_\mathrm{ensem}=10^8$ profile the smoothest and lower-resolution runs exhibiting statistical oscillations in the high-density inner core.
This bell-shaped radial temperature distribution is consistent with the profiles reported by Schanz et al.~\cite{dabh10} and the center-to-wall gradient described by Kim et al.~\cite{kim2007molecular} for different gas--liquid systems, confirming it as a consistent feature of MD sonoluminescent collapse.
The density profile (panel~c) shows a nearly uniform elevated interior with a density enhancement near the bubble wall: during the final collapse stage the inward wall velocity substantially exceeds the thermal diffusion speed of the outer gas shell, so particles accumulate at the moving boundary faster than they can redistribute, producing a wall-adjacent density pile-up consistent with the advection-dominated compression regime.
This wall density enhancement is reported in prior MD studies~\cite{kim2007molecular,ruuth2002molecular} and is more pronounced for $N_\mathrm{ensem}=10^6$ because the larger EP diameter amplifies the per-particle contribution to the outer bins, a coarsening effect that diminishes as $N_\mathrm{ensem}$ increases.
The pressure profile (panel~b) decreases monotonically from center to wall at $t_\mathrm{min}$, with the interior pressure approximately twice the wall-region value for the best-resolved run; the gradient is more pronounced at higher $N_\mathrm{ensem}$, where reduced statistical noise reveals finer radial pressure structure.
This center-to-wall gradient is consistent with the non-uniform pressure profile predicted by the Kwak model (Eq.~\eqref{eq:density}), in which the inertial correction during collapse introduces a quadratic $r^2$ dependence that reduces the pressure toward the bubble wall.

\paragraph{Emission pulse}
Figure~\ref{fig:general-emission} shows the total spatially integrated Saha--LTE emission $L_\mathrm{tot}(t)$ for all three ensemble sizes at fixed full-range ionization and $\alpha_t=1.0$, alongside a normalised comparison and the radial emission distribution at peak emission time.
In all three cases the emission pulse peaks before minimum radius: the peak occurs $0.221\,\text{ns}$, $0.315\,\text{ns}$, and $0.440\,\text{ns}$ before $t_\mathrm{min}$ for $N_\mathrm{ensem} = 10^6$, $10^7$, and $10^8$ respectively.
The peak shifts further before $t_\mathrm{min}$ as $N_\mathrm{ensem}$ increases, consistent with the finer spatial resolution of the inner core resolving the early onset of hot-core formation more precisely.
The absolute peak emission increases with $N_\mathrm{ensem}$: $3.45\times10^{-4}\,\text{mW}$, $4.06\times10^{-4}\,\text{mW}$, and $5.28\times10^{-4}\,\text{mW}$ for $10^6$, $10^7$, and $10^8$.
The normalised panel~(b) shows that the pulse shape is consistent across resolutions and the pulse width is narrow, on the order of $0.1\,\text{ns}$, consistent with the $0.070$--$0.170\,\text{ns}$ full-width at half-maximum reported by Schanz et al.~\cite{dabh10} for similar driving conditions.
Panel~(c) shows the radial emission profile at peak emission time: emission is strongly concentrated near the bubble centre, consistent with the central hot-core structure seen in panel~(a) of Fig.~\ref{fig:scaling-radial} and with the spatially resolved emission profiles of Schanz et al.~(their Fig.~13).

Kim et al.~\cite{kim2007molecular} study a xenon bubble in sulfuric acid with equilibrium radius $R_0 = 15\,\mu\text{m}$ and driving frequency $f = 37.8\,\text{kHz}$---a different gas species, larger equilibrium radius, and acidic solvent. Even so, the same qualitative signature appears in their results: temperature peaks before minimum radius, a hot core is present in the radial profile, and the emission pulse is short relative to the acoustic period.
These commonalities across hard-sphere EBMD~\cite{dabh10,kim2007molecular}, soft-sphere TBMD (this work), and the shared Kwak continuum reference confirm that the essential mesoscopic physics of sonoluminescent collapse is robust to the choice of molecular interaction model and simulation methodology.

\FloatBarrier
\section{Convergence and Computational Performance}
\label{sec:convergence}

Two aspects of the numerical implementation warrant examination beyond the physics results: the accuracy of the time integrator as a function of LJ cutoff radius, and the computational cost of the production ensemble sizes.

\subsection{Integrator convergence}

LAMMPS employs the velocity--Verlet time integration scheme \cite{swope1982computer,verlet1967computer}, which achieves global second-order accuracy in time when the pair potential is at least $C^1$ continuous \cite{hairer2003geometric}.
When a truncated Lennard--Jones potential is used with a short cutoff, however, the force exhibits a discontinuity at $r = r_c$, breaking this continuity and degrading the formal convergence order of the integrator.

\begin{figure}[htbp]
  \centering
  \subfloat[]
  {\includegraphics[width=0.45\textwidth]{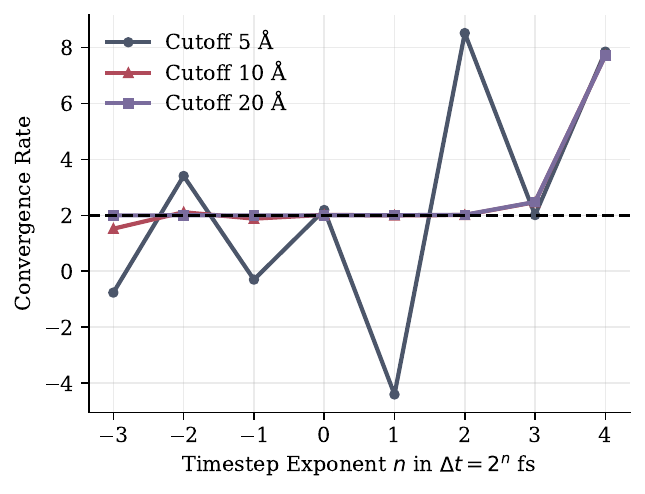}\label{fig:convergence-cutoff-order}}
  \subfloat[]
  {\includegraphics[width=0.45\textwidth]{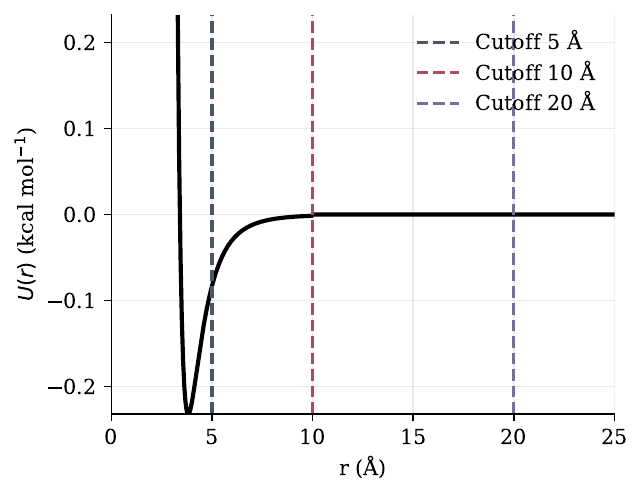}\label{fig:convergence-cutoff-potential}}
  \caption{Integrator convergence as a function of LJ cutoff radius.
  Colors encode cutoff value consistently across both panels:
  gray ($5\,\text{\AA}$), maroon ($10\,\text{\AA}$), purple ($20\,\text{\AA}$).
  (a)~Measured convergence order of the velocity--Verlet scheme as the timestep $\Delta t = 2^n\,\mathrm{fs}$ is refined.
  The dashed horizontal line marks the expected second-order rate.
  Cutoffs of $10$ and $20\,\text{\AA}$ track the reference closely; the $5\,\text{\AA}$ cutoff produces erratic, non-convergent behavior driven by discontinuous forces at truncation.
  (b)~Full LJ potential $U(r)$ (black solid curve) with vertical dashed lines marking the three cutoff positions.
  At $5\,\text{\AA}$ the potential is truncated on the repulsive shoulder where the force is large and discontinuous; at $10$ and $20\,\text{\AA}$ the tail is nearly flat, making the truncation effectively smooth and preserving integrator accuracy.}
  \label{fig:convergence-study}
\end{figure}

To quantify this effect, we designed a controlled 12-particle test system in which particles are placed symmetrically on the coordinate planes and directed toward the origin with equal speed components.
This configuration isolates pairwise LJ forces from boundary and many-body artifacts, providing a clean measurement of integrator accuracy.
Figure~\ref{fig:convergence-study} shows the measured convergence order as the timestep is refined for several LJ cutoff values.
With cutoffs in the range $10$--$20\,\text{\AA}$, the scheme recovers the expected second-order scaling; with shorter cutoffs the rate degrades substantially, confirming that loss of force continuity directly reduces integrator accuracy.
This behavior is consistent with the potential shapes shown in panel~(b): at short cutoffs the interaction energy drops abruptly at $r_c$, whereas at larger cutoffs the tail is nearly flat and the truncation is effectively smooth.

These results justify the cutoff choice used in the large-scale SL simulations.
Although computational constraints preclude arbitrarily large cutoffs at $N_\mathrm{ensem} = 10^8$, an excessively short cutoff would bias both thermodynamic observables and integrator accuracy.
The selected value $r_c \approx 2.15\,\sigma$ is large enough that the truncated LJ interaction is smooth in the dynamically relevant separation range, preserving near-second-order integrator behavior, while keeping the per-timestep force evaluation tractable.

\subsection{Computational performance}

\begin{figure}[t]
  \centering
  \includegraphics[width=0.95\linewidth]{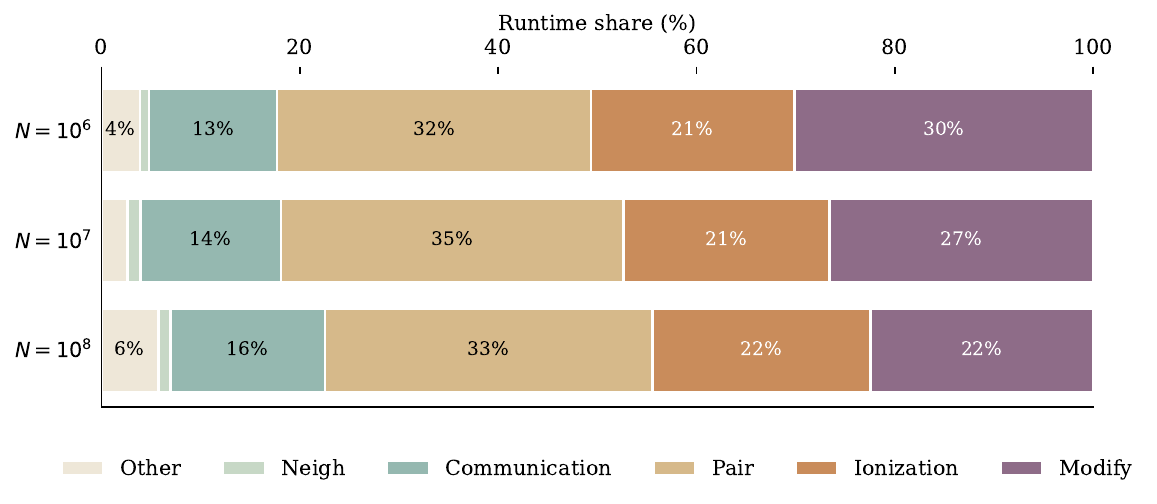}
  \caption{LAMMPS runtime breakdown by component for $N_\mathrm{ensem} = 10^6$, $10^7$, and $10^8$. Pair-force evaluation (LJ + DSF Coulomb) and ionization together account for approximately 55\% of total runtime at all scales; the ionization share remains at 21--22\% across three orders of magnitude in $N_\mathrm{ensem}$, reflecting the cancellation between the enlarged EP reactive radius at small $N_\mathrm{ensem}$ and the reduced absolute particle count. The Modify category comprises the Keller--Miksis wall-coupling update and auxiliary diagnostics.}
  \label{fig:perf_breakdown}
\end{figure}

The three production ensemble sizes span nearly three orders of magnitude in CPU-hours: $N_\mathrm{ensem} = 10^6$ requires 983 CPU-hours on 128 cores, while the $10^7$ and $10^8$ runs each use 2,048 cores and require 55,600 and 975,000 CPU-hours, respectively.
Figure~\ref{fig:perf_breakdown} shows how runtime is distributed across LAMMPS components at all three scales.
The distribution is strikingly stable: pair-force evaluation (LJ + DSF Coulomb) and ionization together consume just over half the total compute time at every scale, with the ionization share holding at 21--22\% across three orders of magnitude in $N_\mathrm{ensem}$.
This invariance reflects two competing effects: as $N_\mathrm{ensem}$ decreases, the enlarged EP reactive radius $r'_\mathrm{ion} = g^{1/2}r_\mathrm{ion}$ increases ionization neighbor-list work per particle, while the reduced absolute particle count offsets this cost in equal measure.

\begin{figure}[!t]
  \centering
  \subfloat[]{
    \includegraphics[width=0.49\linewidth]{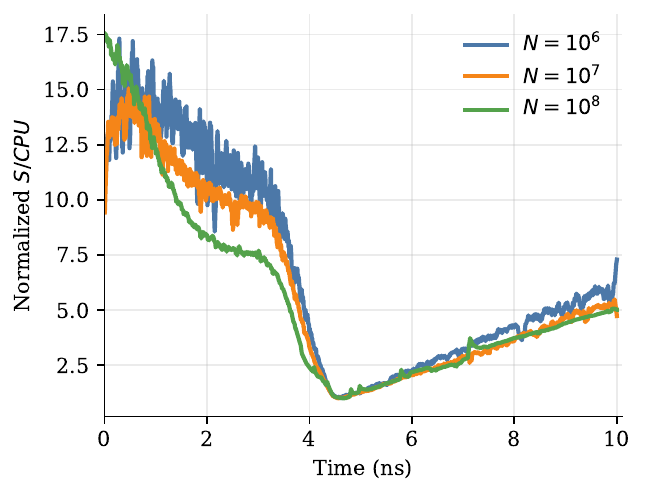}
    \label{fig:perf_throughput}
  }
  \subfloat[]{
    \includegraphics[width=0.49\linewidth]{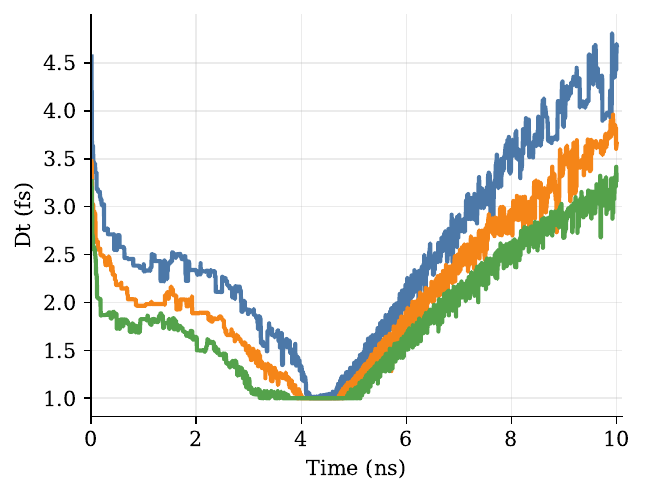}
    \label{fig:perf_dt}
  }
  \caption{Temporal variation of computational cost during the 10-ns simulation window for $N_\mathrm{ensem} = 10^6$ (blue), $10^7$ (orange), and $10^8$ (green).
  (a)~Normalized step throughput ($S/\mathrm{CPU}$), each curve normalized by its own collapse-window minimum. Throughput degrades by approximately 15-fold from the pre-collapse baseline to the minimum near $t \approx 4.2$--$4.7\,\mathrm{ns}$, driven by particle concentration near the bubble center, and recovers symmetrically after the rebound.
  (b)~Adaptive MD timestep $\Delta t$. The timestep contracts to its minimum of ${\sim}1\,\mathrm{fs}$ at the collapse and recovers afterward; larger $N_\mathrm{ensem}$ runs use systematically smaller timesteps because the reduced EP diameter $\sigma' = g^{1/3}\sigma$ tightens the displacement-based step-size criterion. Together, panels (a) and (b) show that the computational cost per nanosecond of simulation time is simultaneously elevated near collapse by both increased per-step work and a reduced timestep.}
  \label{fig:perf_time}
\end{figure}

The temporal distribution of computational work reveals a further structure.
Figure~\ref{fig:perf_time} plots the normalized step throughput and adaptive timestep over the 10-ns simulation window for all three ensemble sizes.
Throughput is approximately 15-fold higher during the pre-collapse phase than at the collapse minimum near $t \approx 4.2$--$4.7\,\mathrm{ns}$: as the bubble radius decreases, particles concentrate near the center and sharply increase both pair-force and ionization neighbor-list work per timestep.
The symmetric recovery after the rebound confirms that the bottleneck is geometric in origin, tied directly to the collapse dynamics rather than to any cumulative state change such as a growing ionization charge distribution.

\section{Conclusions and Future Work}
\label{sec:conclusions}

This work has developed a time-based molecular dynamics framework for simulating single-bubble sonoluminescence within a hybrid continuum--MD formulation.
Unlike prior event-based codes, which are restricted to hard-sphere interactions and event-driven time advancement, the present approach integrates a continuous Lennard--Jones plus damped-shifted-force Coulomb potential at each timestep, enabling self-consistent treatment of ionization state and long-range electrostatics throughout the collapse.
To make such simulations tractable at physical particle counts, an ensemble-particle (EP) scaling formalism maps $N_\mathrm{real}\sim 10^{10}$ physical atoms onto up to $N_\mathrm{ensem} = 10^8$ simulated EPs while preserving temperature, pressure, and ionization statistics, representing a tenfold increase over the largest ensembles reported in prior EBMD studies.
Gas dynamics couple bidirectionally to a continuum Keller--Miksis solver at each timestep, and the resulting predictions are benchmarked against the EBMD reference dataset of Schanz et al.~\cite{dabh10} under identical driving conditions.

The parameter sweep over ionization model and thermal accommodation yields several clear conclusions.
Ionization is by far the dominant regulator of peak temperature: removing the ionization energy sink raises $T_\mathrm{max}$ from the physically plausible range of $17{,}272$--$21{,}265\,\mathrm{K}$, which brackets the Schanz et al.\ reference of $18{,}870\,\mathrm{K}$, to $39{,}617$--$45{,}820\,\mathrm{K}$, a factor-of-two inflation that persists regardless of the electrostatic model chosen.
The thermal accommodation coefficient $\alpha_t$, by contrast, primarily governs the spatially averaged temperature at collapse with limited influence on the peak local compression temperature, a finding consistent with the thermal boundary layer being thin relative to the bubble radius at the collapse minimum.
Scalar observables show consistent, systematic behavior as $N_\mathrm{ensem}$ increases from $10^6$ to $10^8$, validating the EP scaling formalism and confirming that the $10^8$ configuration provides the most reliable reference for all reported quantities.

Several directions warrant future investigation.
The most direct is to eliminate the EP approximation altogether: with the scaling formalism validated at $N_\mathrm{ensem} = 10^8$, a direct $g = 1$ simulation at the full physical particle count (${\sim}10^{10}$) would eliminate modeling and approximation errors due to the use of EPs and provide a numerical ground truth against which the EP results can be assessed.
A second direction is the extension to reactive gas mixtures, since the present model is restricted to monatomic argon; incorporating water vapor, dissolved noble-gas impurities, or nitrogen would require a reactive force field or a QM/MM description of bond-breaking chemistry, and would enable direct comparison with spectroscopic observations of molecular emission during the collapse pulse.
Another interesting direction would be to consider other fluids and fluid models surrounding the sonoluminescing bubble, such as compressible fluids modeled with more complex equations of state \citep{krimans2023power}.
Finally, replacing the empirical Lennard--Jones potential with a machine-learning interatomic potential (MLIP) trained on \emph{ab initio} data at SL-relevant conditions would remove the assumption that LJ parameters calibrated at ambient conditions remain valid at the extreme temperatures and pressures reached during the collapse.
%now that the TBMD framework is established, such a substitution is a natural next step toward a fully first-principles description of the collapse.

\section*{Acknowledgments}

This work was funded in part by an ORAU Ralph E.\ Powe Junior Faculty Enhancement Award.
The authors were supported by NSF Grant No.\ 2324735.
This work used Anvil at Purdue University through allocation PHY250136 from the Advanced Cyberinfrastructure Coordination Ecosystem: Services \& Support (ACCESS) program \cite{10.1145/3569951.3597559}, which is supported by U.S.\ National Science Foundation grants \#2138259, \#2138286, \#2138307, \#2137603, and \#2138296.
D.H.\ thanks Seth Putterman at UCLA for very helpful conversation about sonoluminescence at the beginning of this project.
S.C.\ thanks Ho-Young Kwak at Chung-Ang University for many generous and insightful emails about sonoluminescence.

\bibliographystyle{plainnat}
\bibliography{paper}

\newpage

\appendix
%% ======================================================================
\section{Supplementary Space--Time Heatmaps}
\label{sec:appendix_heatmaps}

The following nine panels provide supplementary space--time views supplementing the results of Sections~\ref{sec:results_scale}--\ref{sec:results_boundary}.

\begin{figure}[!h]
  \centering
  %% Row 1: N-sweep temperature heatmaps
  \subfloat[$N_{\mathrm{ensem}} = 10^6$]{%
    \includegraphics[width=0.32\textwidth]{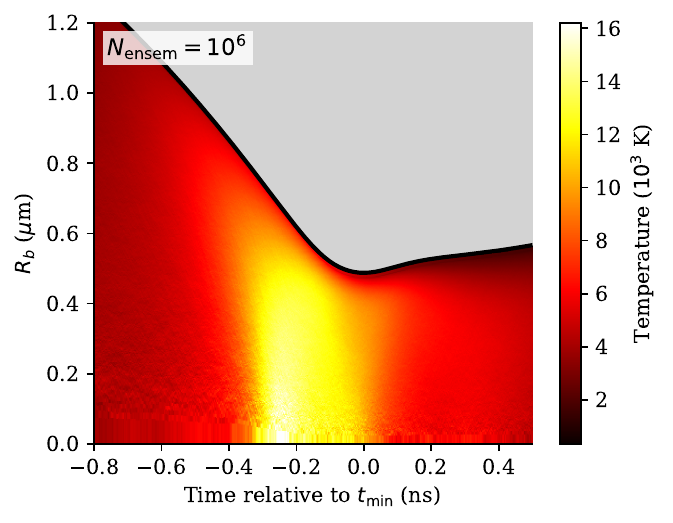}%
    \label{fig:A1-heatmap-N1e6}}
  \subfloat[$N_{\mathrm{ensem}} = 10^7$]{%
    \includegraphics[width=0.32\textwidth]{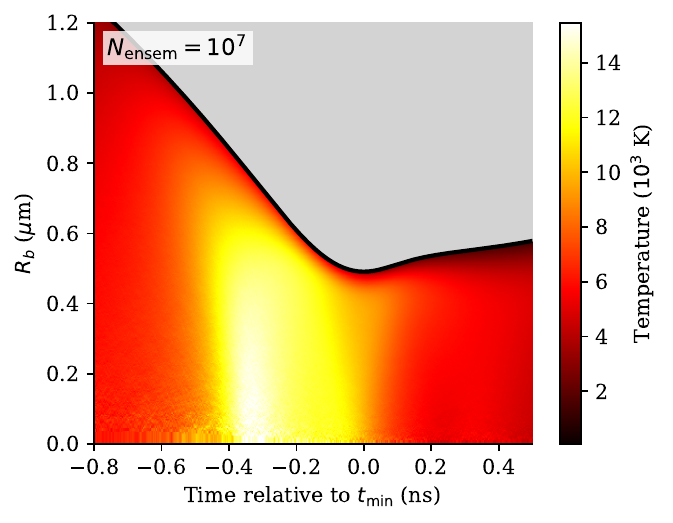}%
    \label{fig:A1-heatmap-N1e7}}
  \subfloat[$N_{\mathrm{ensem}} = 10^8$]{%
    \includegraphics[width=0.32\textwidth]{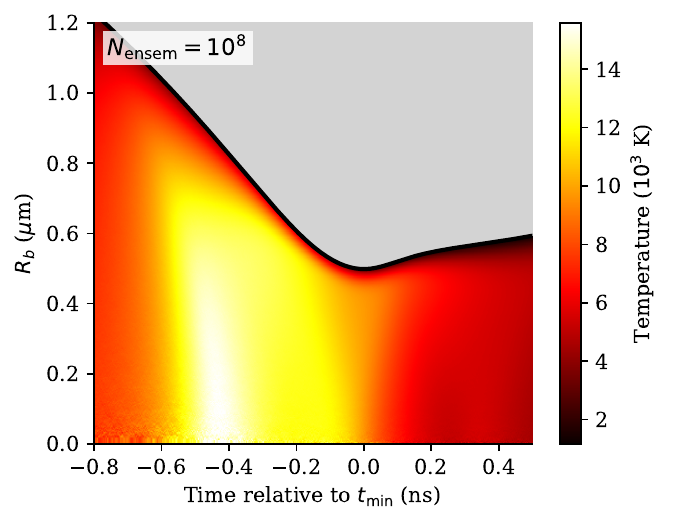}%
    \label{fig:A1-heatmap-N1e8}}
  \\
  %% Row 2: pair-interaction model temperature heatmaps
  \subfloat[No ionization (LJ only)]{%
    \includegraphics[width=0.32\textwidth]{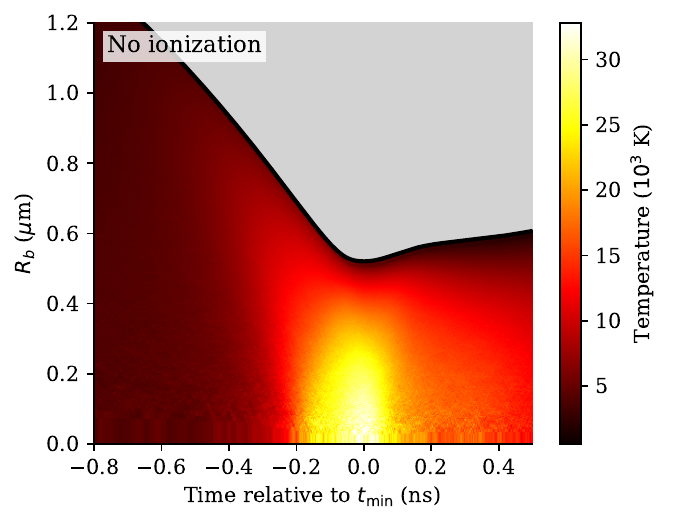}%
    \label{fig:A2-heatmap-noio}}
  \subfloat[Short-range ionization]{%
    \includegraphics[width=0.32\textwidth]{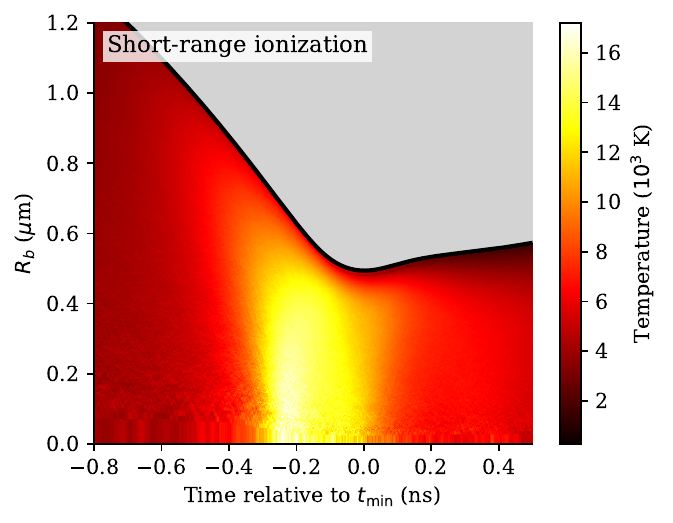}%
    \label{fig:A2-heatmap-shortio}}
  \subfloat[Full-range ionization + DSF]{%
    \includegraphics[width=0.32\textwidth]{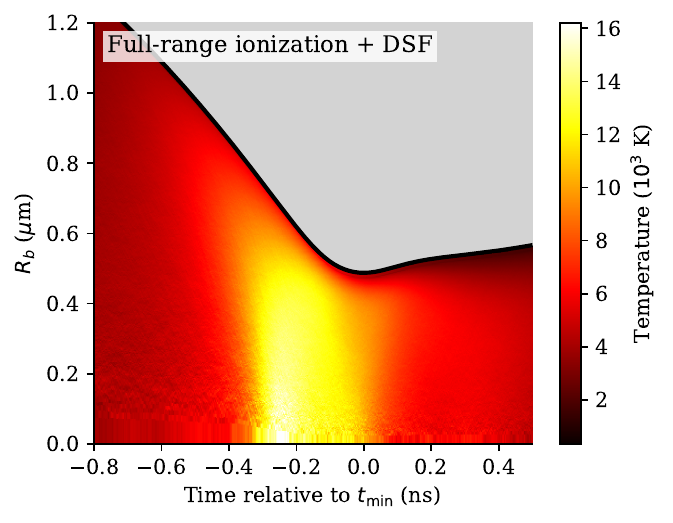}%
    \label{fig:A2-heatmap-fullio}}
  \\
  %% Row 3: emission heatmaps for alpha sweep
  \subfloat[$\alpha_t = 0.0$]{%
    \includegraphics[width=0.32\textwidth]{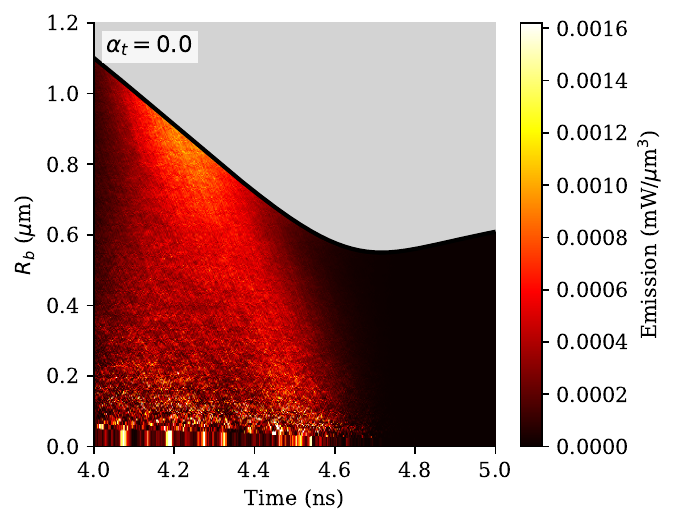}%
    \label{fig:A3-emission-alpha0}}
  \subfloat[$\alpha_t = 0.5$]{%
    \includegraphics[width=0.32\textwidth]{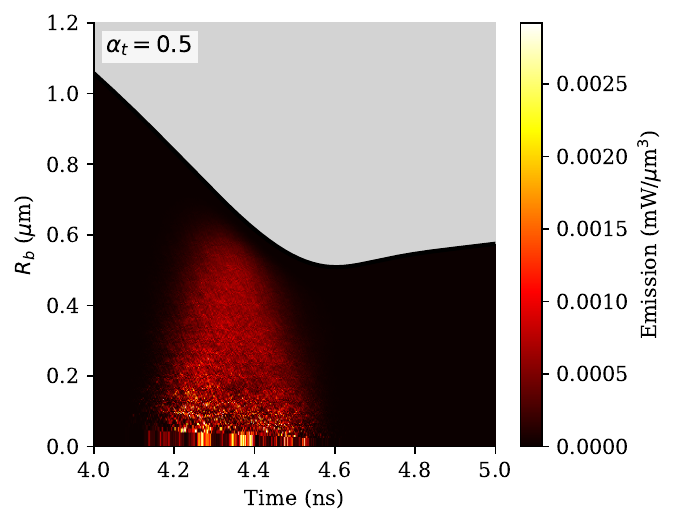}%
    \label{fig:A3-emission-alpha05}}
  \subfloat[$\alpha_t = 1.0$]{%
    \includegraphics[width=0.32\textwidth]{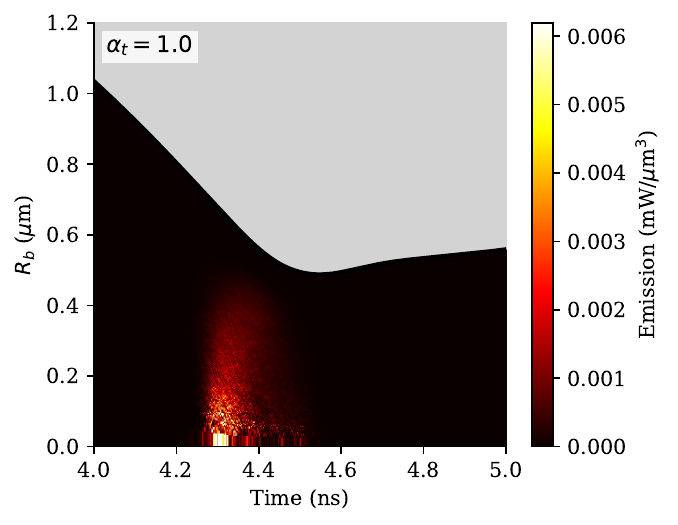}%
    \label{fig:A3-emission-alpha1}}
  \caption{Supplementary space--time heatmaps.
    In all panels the horizontal axis is time, the vertical axis is absolute
    bubble radius $R_b$ ($\mu$m), the solid black curve marks the bubble wall,
    the grey region is the liquid exterior, and each panel uses an independent
    robust display cap at the 99.9th percentile of occupied bins.
    \emph{Row~1} (panels a--c): temperature heatmaps (color in units of
    $10^3\,\text{K}$) for $N_\mathrm{ensem} = 10^6$, $10^7$, and $10^8$ with
    full-range ionization plus DSF Coulomb and $\alpha_t = 1.0$; horizontal
    axis is time relative to $t_{\min}$ (window $[-0.8, 0.5]\,\text{ns}$).
    These panels supplement the radial-profile snapshots of
    Section~\ref{sec:results_scale} by showing the full temporal evolution of
    the hot-core structure across the $N_{\mathrm{ensem}}$ sweep.
    \emph{Row~2} (panels d--f): temperature heatmaps for no ionization
    (LJ only), short-range ionization, and full-range ionization plus DSF
    Coulomb, at $N_\mathrm{ensem} = 10^6$ and $\alpha_t = 1.0$; horizontal
    axis as in row~1.
    These panels connect the emission suppression caused by ionization
    (Section~\ref{sec:results_pair}) to changes in the underlying temperature
    field.
    \emph{Row~3} (panels g--i): Saha--LTE bolometric emission heatmaps
    ($\lambda \geq 200\,\text{nm}$) for $\alpha_t = 0.0$, $0.5$, and $1.0$,
    with $N_\mathrm{ensem} = 10^6$ and full-range ionization plus DSF Coulomb;
    horizontal axis is absolute simulation time (window $[4.0, 5.0]\,\text{ns}$).
    These panels complement the temperature heatmaps of
    Section~\ref{sec:results_boundary} by showing how the spatial
    concentration of the emission pulse evolves with $\alpha_t$.}
  \label{fig:A-heatmaps}
\end{figure}

%% ======================================================================
\clearpage
\section{Radial Profiles Across Parameter Sweeps}
\label{sec:appendix_radial}
%% ======================================================================

The following nine panels provide supplementary radial-profile views across all three parameter sweeps in a unified format.

%% --- Fig B: 3x3 grid --------------------------------------------------
\begin{figure}[!h]
  \centering
  %% Row 1: alpha sweep
  \subfloat[Temperature, $\alpha_t$ sweep]{%
    \includegraphics[width=0.32\textwidth]{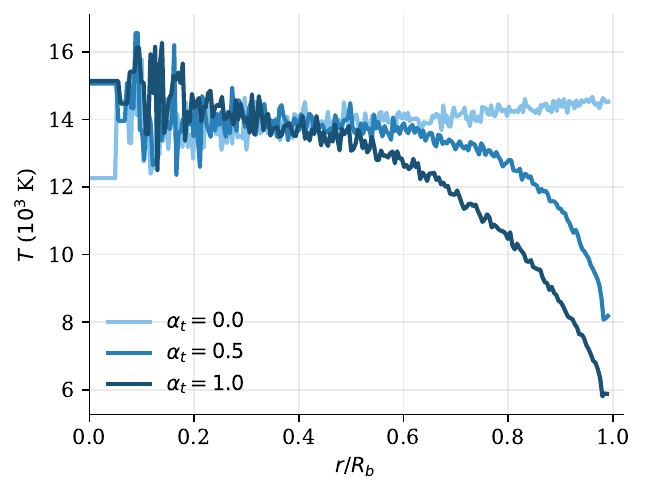}%
    \label{fig:B1-radial-T}}
  \subfloat[Pressure, $\alpha_t$ sweep]{%
    \includegraphics[width=0.32\textwidth]{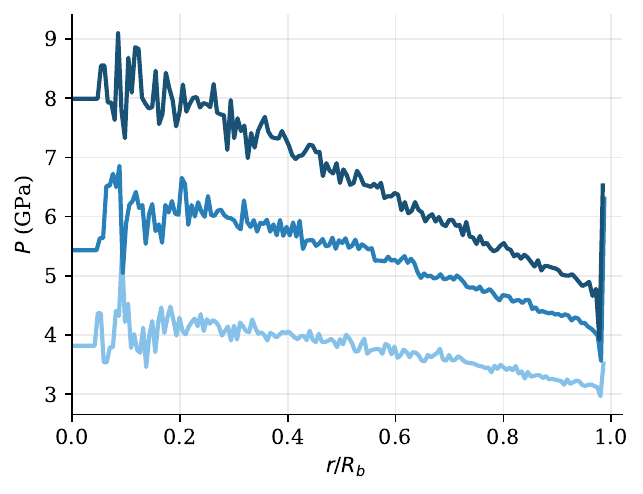}%
    \label{fig:B1-radial-P}}
  \subfloat[Density, $\alpha_t$ sweep]{%
    \includegraphics[width=0.32\textwidth]{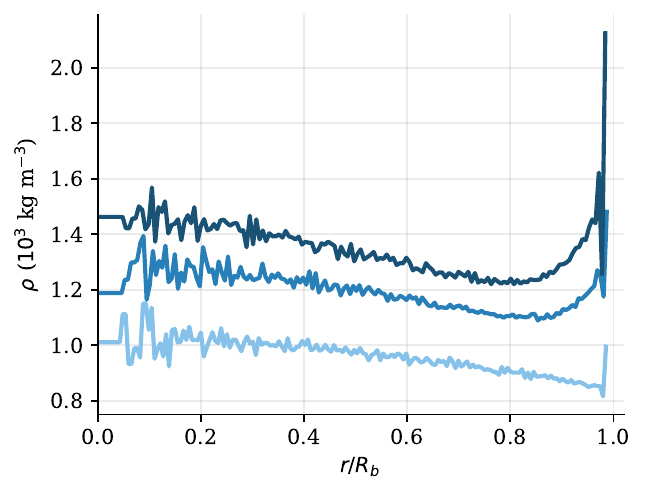}%
    \label{fig:B1-radial-rho}}
  \\
  %% Row 2: pair sweep
  \subfloat[Temperature, pair sweep]{%
    \includegraphics[width=0.32\textwidth]{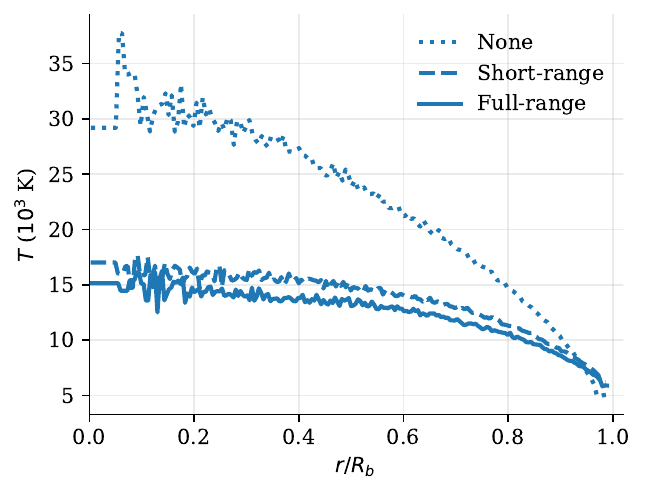}%
    \label{fig:B2-radial-T}}
  \subfloat[Pressure, pair sweep]{%
    \includegraphics[width=0.32\textwidth]{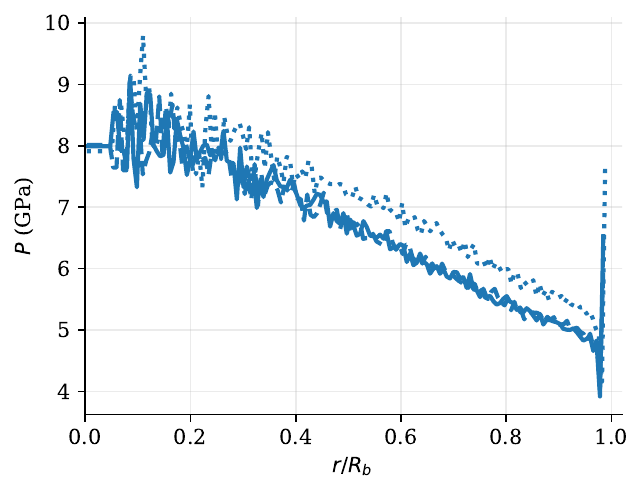}%
    \label{fig:B2-radial-P}}
  \subfloat[Density, pair sweep]{%
    \includegraphics[width=0.32\textwidth]{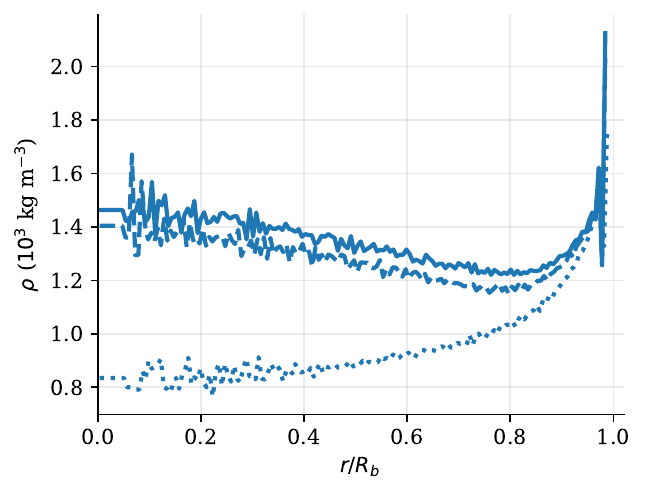}%
    \label{fig:B2-radial-rho}}
  \\
  %% Row 3: N sweep
  \subfloat[Temperature, $N_\mathrm{ensem}$ sweep]{%
    \includegraphics[width=0.32\textwidth]{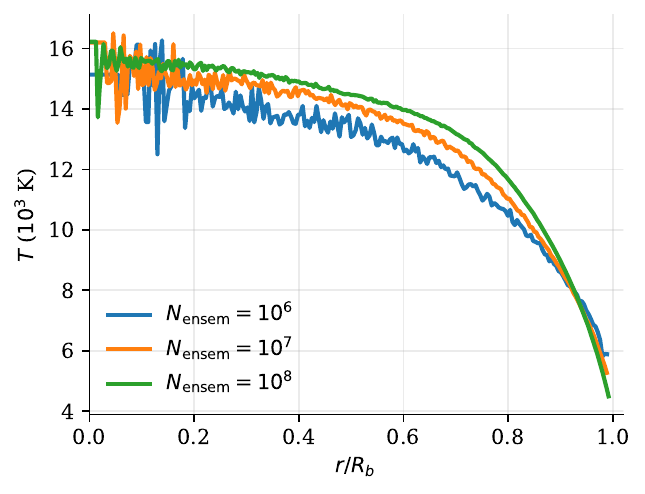}%
    \label{fig:B3-radial-T}}
  \subfloat[Pressure, $N_\mathrm{ensem}$ sweep]{%
    \includegraphics[width=0.32\textwidth]{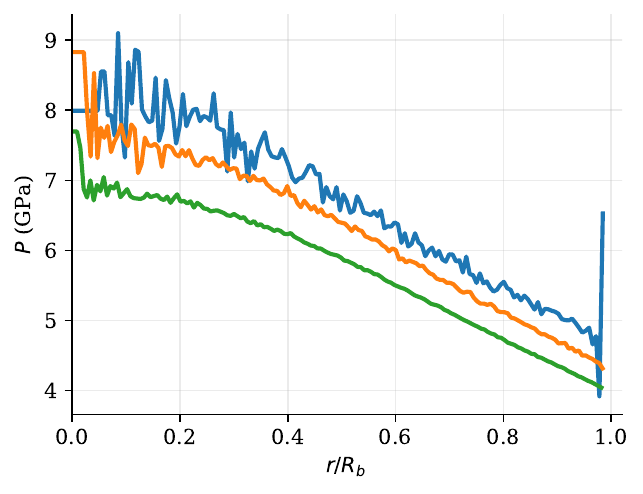}%
    \label{fig:B3-radial-P}}
  \subfloat[Density, $N_\mathrm{ensem}$ sweep]{%
    \includegraphics[width=0.32\textwidth]{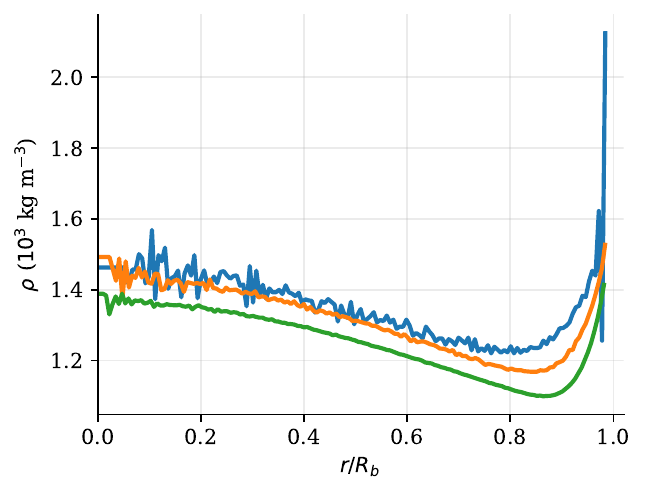}%
    \label{fig:B3-radial-rho}}
  \caption{Supplementary radial profiles at collapse across all three parameter sweeps.
    In all panels the horizontal axis is fractional radius $r/R_b \in [0,1]$
    (outermost two bins excluded); temperature (left column) is evaluated at
    $t(T_{\mathrm{center,max}})$, pressure and density (centre and right columns)
    at $t_{\min}$.
    \emph{Row~1} (panels a--c): $\alpha_t$ sweep at $N_\mathrm{ensem}=10^6$,
    full-range ionization plus DSF Coulomb.
    Light blue: $\alpha_t=0.0$ (adiabatic); medium blue: $\alpha_t=0.5$;
    dark navy: $\alpha_t=1.0$ (isothermal).
    These profiles supplement Section~\ref{sec:results_boundary}, showing how
    wall coupling modifies the radial shape of the hot core and the interior
    pressure uniformity.
    \emph{Row~2} (panels d--f): pair-interaction sweep at $N_\mathrm{ensem}=10^6$,
    $\alpha_t=1.0$.
    Blue solid: full-range ionization plus DSF; blue dashed: short-range
    ionization; blue dotted: no ionization (LJ only).
    Panel~(d) uses a shared temperature axis to make the factor-of-two
    suppression from ionization directly apparent.
    These profiles supplement Section~\ref{sec:results_pair}.
    \emph{Row~3} (panels g--i): $N_\mathrm{ensem}$ sweep at $\alpha_t=1.0$,
    full-range ionization plus DSF Coulomb.
    Blue: $N_\mathrm{ensem}=10^6$; orange: $10^7$; green: $10^8$.
    These panels reproduce the data of Fig.~\ref{fig:scaling-radial} in a
    format consistent with the other rows of this figure, showing the
    resolution dependence of the radial structure at collapse.}
  \label{fig:B-radial-sweeps}
\end{figure}

%% ======================================================================
\clearpage
\section{Scalar Parameter Sweep Summary}
\label{sec:appendix_scalar}
%% ======================================================================

The following six panels visualise all peak scalar observables across the full eleven-configuration parameter sweep (Table~\ref{tab:sweep}).

%% --- Fig C.1 ----------------------------------------------------------
\begin{figure}[!h]
  \centering
  \subfloat[Peak local temperature $T_{\max}$]{%
    \includegraphics[width=0.48\textwidth]{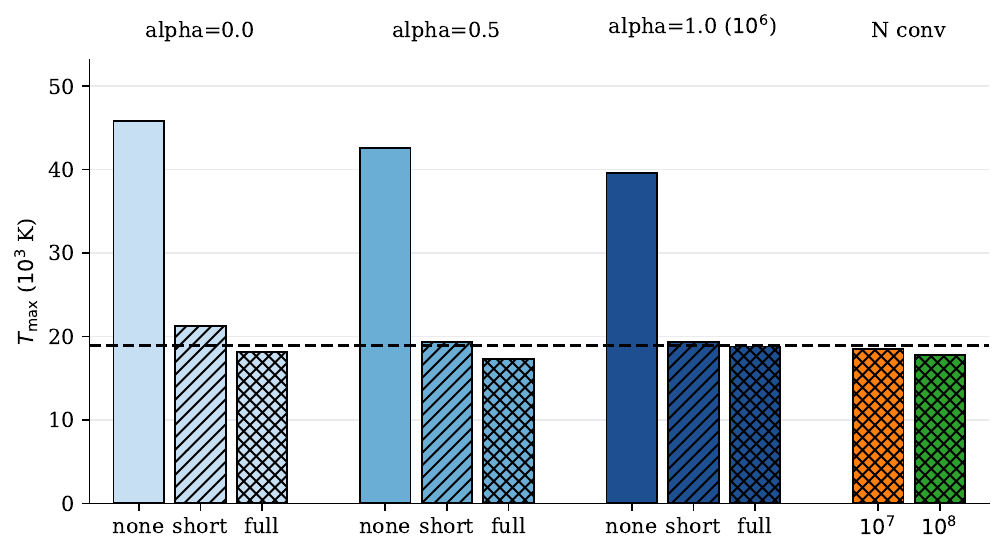}%
    \label{fig:C1a-Tmax}}
  \subfloat[Peak volume-averaged temperature $T_{\mathrm{av,max}}$]{%
    \includegraphics[width=0.48\textwidth]{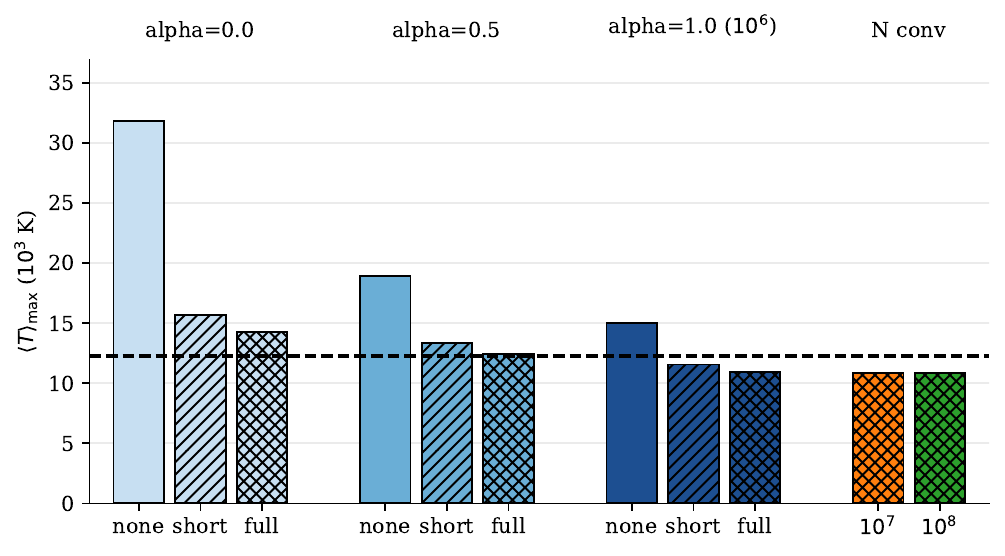}%
    \label{fig:C1b-Tavmax}}
  \\
  \subfloat[Cumulative ionization fraction $f_{\mathrm{ion}}$]{%
    \includegraphics[width=0.48\textwidth]{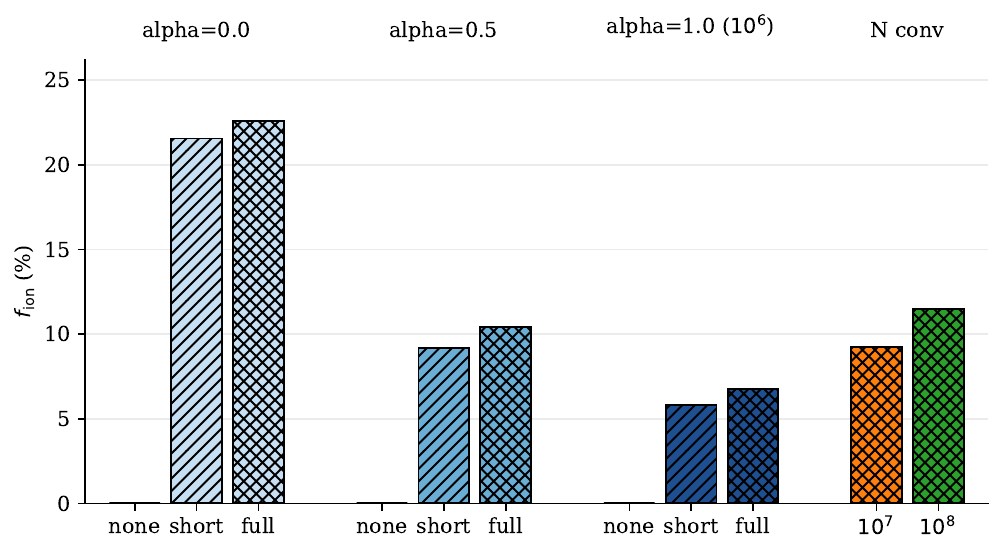}%
    \label{fig:C1c-fion}}
  \subfloat[Minimum bubble radius $R_{\min}$]{%
    \includegraphics[width=0.48\textwidth]{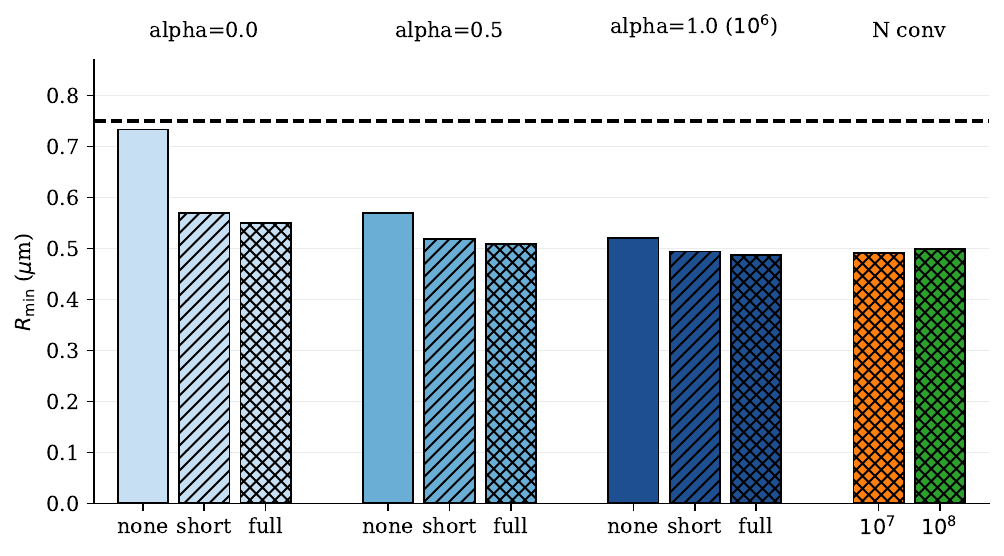}%
    \label{fig:C1d-Rmin}}
  \\
  \subfloat[Maximum wall speed $v_{W,\max}$]{%
    \includegraphics[width=0.48\textwidth]{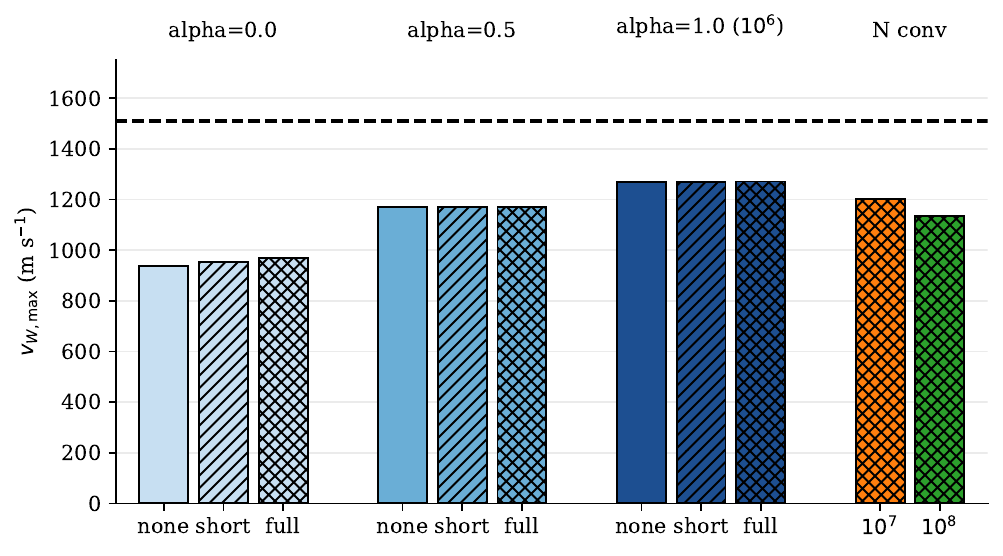}%
    \label{fig:C1e-vWmax}}
  \subfloat[Peak wall pressure $p_{W,\max}$]{%
    \includegraphics[width=0.48\textwidth]{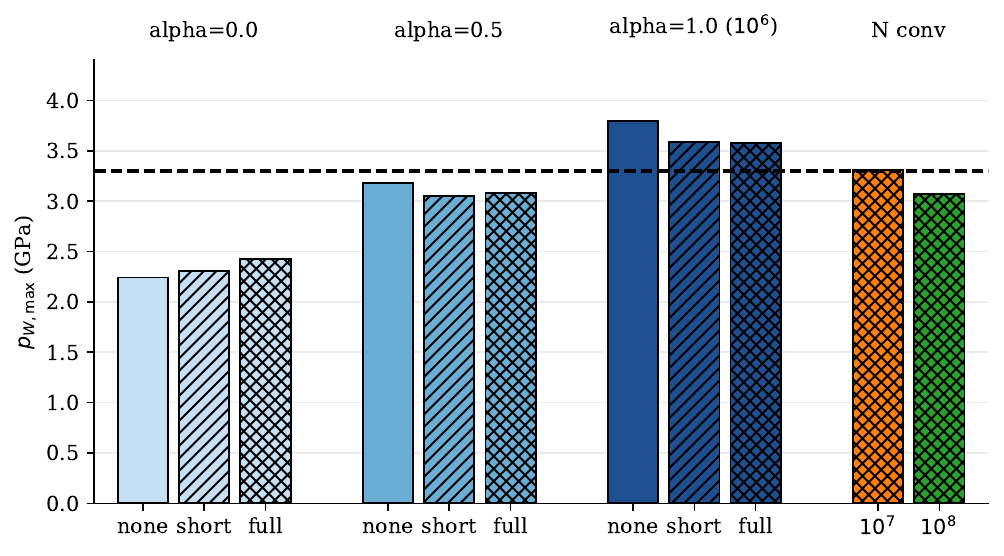}%
    \label{fig:C1f-pWmax}}
  \caption{Peak scalar observables across all 11 simulated configurations
    (Table~\ref{tab:sweep}).
    Bars are grouped into four blocks separated by gaps: three $\alpha_t$ blocks
    for $N_\mathrm{ensem} = 10^6$ (light blue: $\alpha_t = 0.0$; medium blue:
    $\alpha_t = 0.5$; dark navy: $\alpha_t = 1.0$) and one $N$-convergence
    block (orange: $N_\mathrm{ensem} = 10^7$; green: $10^8$, both at
    $\alpha_t = 1.0$ and full-range ionization).
    Within each $\alpha_t$ block, bar fill encodes the pair-interaction model:
    solid fill = no ionization (LJ only); diagonal hatching = short-range
    ionization; crosshatch = full-range ionization plus DSF Coulomb.
    The dashed horizontal line marks the Schanz et al.\ EBMD reference value
    where available~\cite{dabh10}.}
  \label{fig:C1-scalar-sweep}
\end{figure}

\end{document}